# Design of ANF/MXene/SSG sandwich structure with electromagnetic shielding performance and impact resistance

**Author:** Kai Wang [a], Chiyu Zhou [a], Jianbin Qin [a]

[a] Northwestern Polytechnical University, Xi'an 710072, Shaanxi, China.

**Abstract**

Since entering the information era, electronic devices gradually play an important role in people's daily lives. However, the abuse of electronic devices leads to corresponding electromagnetic (EM) wave pollution. The complex external environment causes the potential for physical impact. In this work, an ANF/MXene/SSG flexible sandwich structure was fabricated according to methods of vacuum filtration, directional freeze-casting solidification, and polyurethane encapsulation. This sandwich structure exhibits excellent electromagnetic interference shielding effectiveness (EMI SE) and impact resistance. The ANF/MXene/SSG/MXene/ANF (A/M/S/M/A) structure with a thickness of 4.48 mm, exhibits an EMI SE of 62.6 dB in the X-band. The main EM wave attenuation mechanism is microscopic multireflection and ohmic loss absorption. Simultaneously, this A/M/S/M/A structure can attenuate the highest 71.0% of impact force. Apart from its excellent protection function, the sandwich structure also acts as a human body movement sensor. The PDMS/MGN triboelectric nanogenerator (TENG) fabricated from the A/M/S/M/A structure in this work can generate voltage under external stimuli, achieving the recycle of mechanical energy.

## 1. Introduction

The development of high-end electronic devices and technology has influenced industry and human lives deeply. However, corresponding electromagnetic (EM) radiations have affected human health. Therefore, electromagnetic interference shielding materials with excellent performance are now in great demand [1-3]. Due to the complex working environment, impact resistance and other mechanical properties are also gradually becoming important considerations. Tough metal-based shielding materials which exhibit good electromagnetic interference shielding effectiveness (EMI SE) and anti-impact performance have been developed, however, their high density and poor corrosion resistance have restricted their applications [1, 4]. Nowadays, conductive polymer composites (CPCs) have become options for shielding materials design due to their low density, low cost, higher corrosion resistance, and facilitative structure design and manufacturing [3, 5, 6]. After entering the era of artificial intelligence, wearable devices are gradually entering into people's daily lives. During the rapid development of wearable devices, flexible wearable shielding materials with excellent anti-impact performance are in high demand for all types of people in need of protection [7-9].

Recently, CPCs with different structure designs such as homogenous structures, isolated structures [10], layer structures [11, 12], porous structures [13, 14] or a combination [15-17] have been developed. Due to the synergic performance with composites for either conductivity or mechanical properties [18, 19], nanomaterials such as graphene (including graphene oxide (GO) and reduced graphene oxide (rGO)) [20, 21], carbon nanotubes (CNTs) [20, 22], silver nanowires (AgNWs) [23, 24], and nickel nanowires (NiNWs), or nickel nano chains (NiNCs) [15, 25] have become favoured by researchers in various shielding materials. However, to fabricate CPC in a homogenous structure and isolated structure, percolation value is important, which means adequate fillers are needed to reach critical percolation value [26, 27]. The increase of conductive filler affects the flexibility, toughness, and processibility of CPC, indicating the difficulty of fabricating flexible shielding materials [28, 29]. Among all the different structure designs, layer structure exhibits excellent potential for flexible EMI shielding material to the smallest thickness. The application of layers conductive material such as graphene enhances EMI SE of film composites by multilayer reflection. Wan et al. [30] fabricated lightweight and flexible graphene papers with large-sized graphene sheets (LG), exhibiting EMI SE up to ~52.2 dB at 8.2 GHz with a thickness of 12.5 μm of iodine-doped LG films. Zhang et al. [31] achieved the highest EMI SE of 77 dB by sandwiching 3 layers of graphene films with poly(arylene ether nitrile) (PEN). The composite film also exhibited high flexibility and tensile strength. In addition to carbon series conductive materials, a new kind of 2D conductive material has been getting attention. MXene is a two-dimensional transition metal carbide and/or nitride with the formula $M_{n+1}X_nT_x$ [32, 33]. In this formula, M represents an early transition metal (Ti, Zr, V, Nb, Ta, Mo, etc.), X represents carbon and/or nitrogen, and T represents the surface functional group (-OH, =O, -F, etc.). Metallic conductivity, excellent mechanical properties, and good interaction with organic molecules and ions shows MXene to be a good candidate for CPCs [32-34]. Among about 20 types of MXene, $Ti_3C_2T_x$ is the most commonly studied for MXene/polymer composites [35]. Shahzad et al. [36] prepared $Ti_3C_2T_x$ films with thicknesses of 2.5 μm and 45 μm, exhibiting EMI SE of > 50 dB and 92 dB respectively. Additionally, the research group also applied sodium alginate (SA) to fabricate $Ti_3C_2T_x$-SA composite films, exhibiting a maximum EMI SE of 57 dB for 90 wt% of $Ti_3C_2T_x$ with a thickness of 8 to 9 μm.

In addition to good potential for flexible EM shielding materials, layer structure offers the ability for functional design. The composite can achieve multi-function performance by designing corresponding functional layers. For example, the conductive polymer, polypyrrole (PPy), can be intercalated into the layers of $Ti_3C_2T_x$ to expand interlayer space, facilitating charge transport [37]. Polymer matrices with excellent mechanical properties can also be introduced into flexible shielding layer structures. Liu et al. [38] fabricated $Ti_3C_2T_x$/poly(3,4-ethylenedioxythiophene)–poly(styrenesulfonate) (PEDOT:PSS) composite film for EMI SE shielding with a thickness of 11.1 μm when PEDOT:PSS weight ratio is 7:1. The tensile strength increased from 5.62 to 13.71 MPa compared with pure $Ti_3C_2T_x$ film. Additionally, a nanofiber network has been constructed to enhance the mechanical properties of flexible MXene films in recent years. MXene/cellulose nanofiber composite paper was fabricated by Cao et al. [39] with excellent EMI SE and mechanical properties. According to the interaction of cellulose nanofiber and MXene, the tensile strength of this composite paper is up to 135.4 MPa, and the fracture strain is up to 16.7%.

Apart from cellulose nanofibers, aramid nanofibers (ANFs) based on poly(p-phenylene terephthamide) (PPTA) are getting increasing attention from researchers. ANFs are nanoscale one-dimensional PPTA fibres, with hydrogen bonds, van der Waals forces and π-π stacking to enhance molecular strength. Therefore, ANFs exhibit excellent mechanical properties, thermal stability, and chemical corrosion resistance, hence are often applied as the reinforcement of composites [40, 41]. Tian et al. [42] fabricated ANF-insulated paper through the bottom-up polymerization-induced self-assembly process and vacuum-assisted filtration. Their ANF paper exhibits excellent mechanical strength (up to 60 MPa) and thermal stability. Zhou et al. [43] introduced carboxylated chitosan into the ANF system (C-ANFs) as a hydrogen bonding donor and soft-locking agent, achieving 2.41 times higher tensile strength and 32.69 times higher toughness compared with pure ANF networks. As for flexible shielding films, Ma et al. [29] fabricated ANF-MXene/AgNWs nanocomposite paper via two-step vacuum filtration, exhibiting a maximum EMI SE of ~80 dB when the weight ratio of MXene/AgNWs is 80%. The tensile strength of this double-layered structure reaches 235.9 MPa and fracture strain reaches 24.8%. This research indicates that ANFs are suitable for the panel of MXene flexible shielding layer structure.

Materials with shear stiffening ability can exhibit a reverse response to external stimuli, which can be applied for anti-impact material fabrication. SSG is one type of boron-siloxane polymer network with a dynamic low crosslinking degree [44, 45]. Due to the self-healing ability and stress field adaptability, SSG has been widely applied for impact protection. Compared with shear thickening fluid (STF), SSG avoids the problem of particle sedimentation and liquid volatilization, leading to relatively stable performance [44, 46]. However, the cold flow of SSG under gravity is inevitable, which means the dimensional stability of SSG is poor [47]. Therefore, SSG is usually encapsulated prior to application. Xu et al. [48] fabricated a sandwich structure consisting of two Kevlar layers and a Silly Putty core. The edge of the Kevlar layer was stitched to maintain a soft SSG core inside. All the energy can be dissipated by the sandwich structure when the impact velocity is below 110 m/s. The maximum energy dissipation was 20.8 J, which indicates a 60% increment compared to pure Kevlar. In addition to encapsulating SSG, two common strategies are applied to modify the dimensional stability of SSG. One strategy is to introduce a permanent network that acts as a scaffold for a B-O dynamic network. Elastic-SSG was fabricated by Wang et al. [49] by introducing methyl vinyl silicone rubber (VMQ) corresponding with heat pressing. Elastic-SSG occupies both shear stiffening performance by the remaining B-O dynamic bonding and elastic property of silicone rubber. Wu et al. [50] applied polydimethylsiloxane (PDMS) to interpenetrate SSG, preparing elastic-SSG to overcome fluidity and irreversible deformation of SSG. Another method is to introduce SSG into a porous matrix. Wang et al. [51] applied the "dip and dry" method to immerse the polymer matrix into a polyurethane (PU) sponge and remove acetone by heating the composite under vacuum. It can be shown that the sizes and dimensions of the composite are almost those of the pure polymer matrix with the increasing time. It also exhibits excellent impact and fatigue strength, accompanied by dimensional stability under cyclic utilization. According to existing research, SSG or SSG composite exhibits excellent candidacy as an energy-absorbing core of a multilayer flexible structure.

Apart from protection from EM waves and external impact, some wearable devices also feature in monitoring human activities. In the artificial intelligence era, effective data collection is one of the chief problems for smart wearable devices. Since its invention [52], a triboelectric nanogenerator (TENG) has been used as one device that can scavenge almost any kind of ambient mechanical energy in daily life into electricity [53, 54]. Therefore, a wearable TENG device can act as a self-powered sensor to monitor external stimuli [55]. The basic theory of TENG is triboelectric effect and electrostatic induction [56-58]. Nowadays, various TENGs with protection features have been investigated by researchers. Wang et al. [59] fabricated enhanced Kevlar-based TENG (EK-TENG) with stable sensing capability in harsh loading environments. Under low-speed impact, the 30-layer EK-TENG can dissipate impact force from 1820 N to 439 N. Hou et al. [60] fabricated one kind of flexible one-mode triboelectric nanogenerator with excellent puncture resistance, tear resistance, and self-healing ability. This TENG can be applied to monitor human body movement or detect road surface roughness.

In this work, an ANF/MXene/SSG sandwich structure will be fabricated for EMI shielding and impact resistance. ANFs will be fabricated from aramid short-cut fibres dispersed in dimethylsulfoxide (DMSO). ANF/MXene double layers will be assembled by two-step vacuum filtration, in which ANF is the anti-impact panel and MXene is the conductive layer for EMI shielding. The ANF/MXene double layer exhibits excellent tensile strength and modulus, indicating the improvement compared with the pure MXene single layer. As an energy-absorbing core, SSG is constructed by B-O dynamic crosslinking and immersed into oriented PU foam to achieve dimensional stability. The PU foam is fabricated by directional freeze casting solidification to introduce the elasticity into SSG/PU composite without sacrificing the shear stiffening effect of SSG. After encapsuling by waterborne polyurethane (WPU), this ANF/MXene/SSG sandwich structure exhibits excellent EMI SE and adequate anti-impact ability with flexibility. Apart from protection, this sandwich structure can be converted to a flexible TENG after introducing the GO/Ni system into the conductive layer and adding grooves on the energy-absorbing core to offer more space to separate two electrodes. This TENG exhibits the ability to generate voltage, indicating the potential as a human body monitor for smart wearable devices and artificial intelligence areas. The EMI shielding theory, impact resistance theory, and TENG theory are also discussed thoroughly in this work.

## 2. Experimental

### 2.1 Materials

Dimethylsulfoxide (DMSO), potassium hydroxide (KOH), lithium fluoride (LiF), absolute ethanol, and boric acid were purchased from Guangdong Guanghua Chemical Co., Ltd. China. Hydroxyl silicone oil was purchased from Jinan Xingchi Chemical Co., Ltd. China. Waterborne polyurethane was purchased from Shenzhen Jitian Chemical Co., Ltd. China. Aramid short-cut fibers were purchased from Shenzhen Teli new material technology Co., LTD, China. $Ti_3AlC_2$ (MAX) was purchased from 11 Technology Co., Ltd. hydrochloric acid (HCl) was purchased from Shanghai Hushi

Laboratory Equipment Co., Ltd. GO was purchased from Shaanxi Coal Chemical Industry Technology Research Institute Co. Ltd, China. NiNCs synthesized in previous published work [15] were used. 184 PDMS rubber and curing agent were purchased from Dow Corning, USA. Deionized water used in the experiment was supplied by Hangzhou Wahaha Group Co., Ltd. China. All the chemicals and reagents were used as received.

**2.2 Preparation of ANF dispersion**

ANF dispersion was prepared by dispersing aramid short-cut fiber into KOH/DMSO solution. 4 g aramid short-cut fibers and 6 g KOH were added into 2 L liquid state DMSO. The mixture was stirred at 300 rpm at 35 °C for 7 days to obtain a dark red solution. This solution was diluted using liquid state DMSO and deionized water at a volume ratio of 1:9:1 to fabricate ANF dispersion in light yellow colour (Fig. 1).

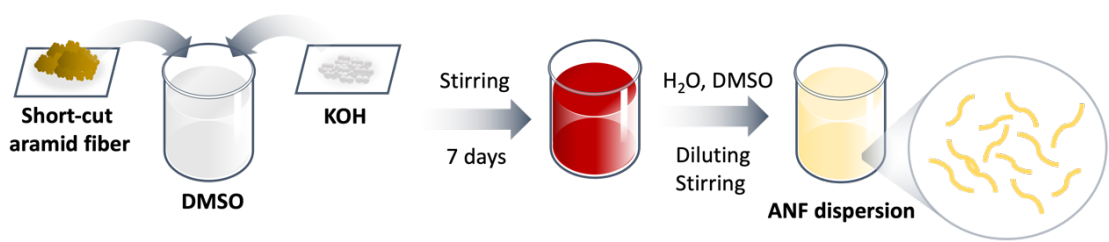

Fig. 1 *Schematic illustration of the preparation of ANF dispersion*

**2.3 Preparation of aqueous MXene**

Aqueous MXene ($Ti_3C_2T_x$) was fabricated by mild etching method. 30 mL of hydrochloric acid and 10 mL of deionized water were mixed. After that, 2 g LiF was added. The mixture was stirred for 5 minutes before 2 g $Ti_3AlC_2$ (MAX) was gradually added to the mixture. The mixture was stirred at 35 °C for 24 h to etch the Al layer from

Ti$_3$AlC$_2$. The product was washed with deionized water repeatedly until the pH was above 6. The black product was treated with ultrasonication in a cell shredder for 1.5 h at 60 % power to get multilayer MXene. After ultrasonication, the dark green liquid product was centrifuged at 3000 rpm for 30 min to collect the upper suspension. The centrifugation was repeated 3 times to get aqueous MXene (Fig. 2).

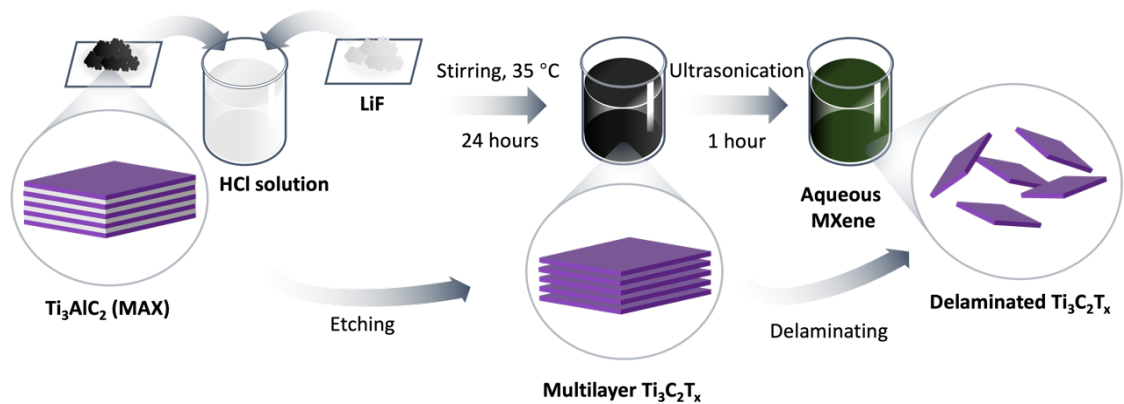

Fig. 2 *Schematic illustration of the preparation of aqueous MXene*

**2.4 Fabrication of ANF/MXene/SSG sandwich structure**

In this work, two types of conductive layers were applied. One kind of conductive layer material was pure MXene, while the other was a MXene/GO/NiNCs homogenous system. The double-layered ANF/MXene or ANF/MXene-GO-NiNCs (ANF/MGN) composite membrane was fabricated by alternative vacuum filtration. The first ANF layer was fabricated by pouring 110 mL ANF dispersion (concentration 0.182mg/mL) for vacuum filtration on a nylon membrane (Vacuum filtration 1). After the formation of the ANF membrane on the nylon membrane, a volume of conductive layer aqueous mixture was poured on the ANF membrane to take on vacuum filtration (Vacuum filtration 2) to form a conductive layer. The double-layered membrane was dried by hanging at room temperature (Fig. 3). In the same way, pure ANF membrane and

MXene membrane were also fabricated. Detailed parameters for each membrane are listed in Table 1.

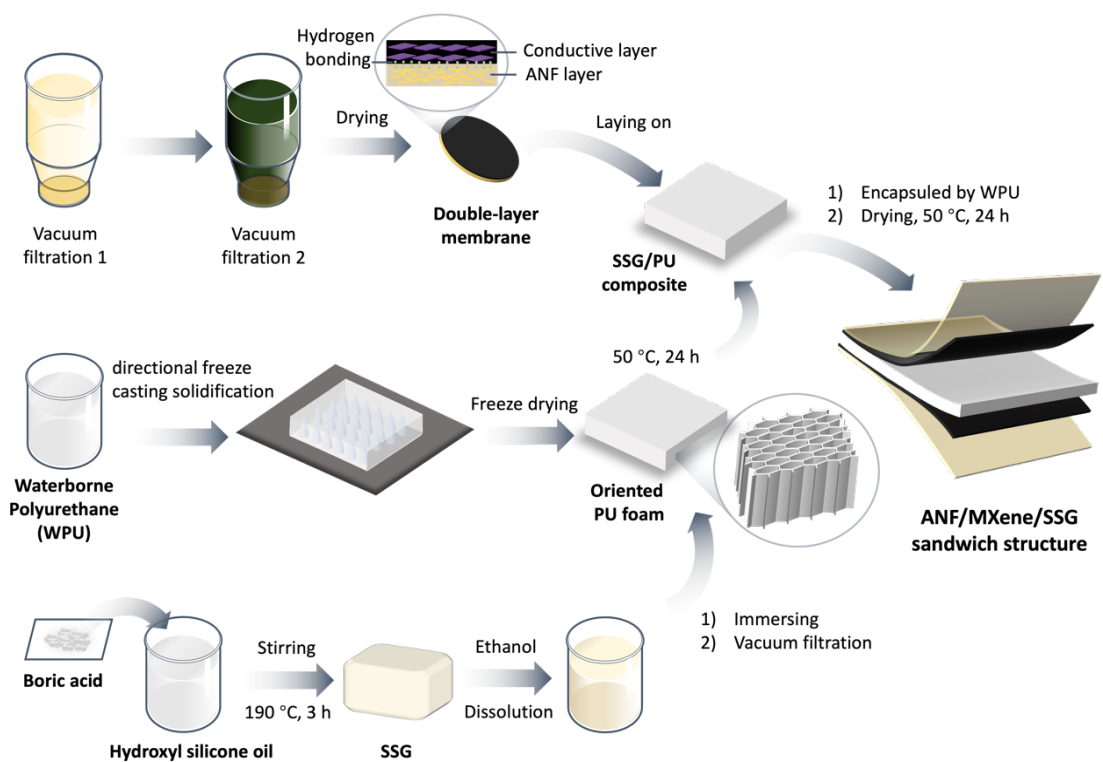

Fig. 3 *Schematic illustration of the fabrication of ANF/MXene/SSG sandwich structure*

Table 1. *Detailed parameter of each membrane used*

| Series | Content (mg) | | | |
|---|---|---|---|---|
| | ANF | MXene | GO | NiNCs |
| Single ANF | 20 | - | - | - |
| Single MXene | - | 40 | - | - |
| ANF/MXene | 20 | 40 | - | - |
| ANF/MGN | 20 | 35 | 5 | 20 |

Hydroxyl silicone oil and boric acid were mixed at the mass ratio of 20:1 and heated at 190 °C for 3 h with stirring. After cooling to room temperature, shear stiffening gel was fabricated. SSG was dissolved in ethanol at a mass ratio of 2:1 to prepare an SSG/ethanol solution. Waterborne polyurethane solution was treated using freeze casting cooled from one end. Oriented ice crystals were produced occupying the space between the polyurethane chains producing orientation. Therefore, oriented holes remained after freeze-drying. SSG/ethanol solution was immersed into oriented PU foam. With the assistance of vacuum filtration, all the holes were filled by the solution. SSG-filled PU foam was heated at 50 °C for 24 h to remove the ethanol. After laying the double-layered composite membrane on the SSG/PU composite, waterborne polyurethane was used to cover the membrane as encapsulation. The composite was treated at 50 °C for 24 h to dry the polyurethane solution, fabricating the ANF/MXene/SSG sandwich structure.

## 2.5 Fabrication of triboelectric nanogenerator

The triboelectric nanogenerator was fabricated by introducing PDMS rubber and copper foil as the cathode and anode, respectively. The uncured PDMS rubber and crosslinker were mixed homogenously at a mass ratio of 10:1. After covering the mixture on both sides of the SSG/PU composite, the PDMS rubber layer was cured at 50 °C for 3 h. Two PDMS hollow layers were applied to cover both sides of the SSG/PU/PDMS composite, forming two grooves on each side. The two grooves offer enough distance between the PDMS rubber layer and the conductive layer to contact and separate, exhibiting better performance. Finally, the triboelectric nanogenerator was fabricated by assembling copper foil, ANF layer, conductive layer, and SSG/PU/PDMS composite. Two copper wires were attached to the copper foil and PDMS layer separately (Fig. 4).

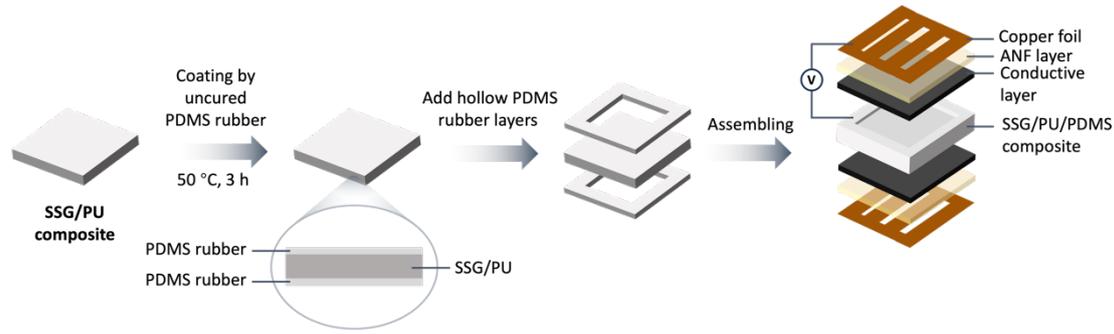

Fig. 4 *Schematic illustration of the fabrication of triboelectric nanogenerator*

**2.6 Characterization**

X-ray diffraction patterns of ANF, MXene, and ANF/MXene membranes were obtained by X-ray diffractometer (XRD, D8 DISCOVER A25) with a Cu Kα source. The character peaks of functional groups and character covalent bonds of ANF, MXene, ANF/MXene, and ANF/MGN membrane were obtained by Fourier Transform Infrared spectroscopy (FI-IR, Bruker Tensor 27). X-ray photoelectron spectrometer (XPS, Kratos Axis Ultra DLD) was used to analyze the surface elements of ANF, MXene, ANF/MXene, and ANF/MGN membrane. The shear stiffening effects of SSG and SSG/PU were characterized by a rotational rheometer (TA Instruments, DHR-20). The micro morphologies of ANF, MXene, ANF/MXene, ANF/MGN membrane, PU foam, and SSG/PU composite were characterized by field emission scanning electron microscope (FE-SEM, FEI Verios). The samples were sputter-coated with a gold layer before SEM observation (SuPro Instruments, ISC 150 Ion Sputter Coater). ANF, MXene, ANF/MXene, and ANF/MGN membranes were cropped into rectangular shape. The electrical conductivity of rectangular membranes was measured using a multimeter (UNI-T, UT58D, 4 significant figures) and calculated using equations (1) and (2).

$$R = \rho(l/S) \quad (1)$$

$$\kappa = 1/\rho \quad (2)$$

Where *R*, *l*, and *S* represent the resistance, length, and cross-section area of the membrane. ρ and κ represent the resistivity and conductivity of membranes.

EMI SE of flexible sandwich structures was characterized by Vector Network Analyzer (VNA, Agilent, N5232A, USA) under the frequency of X-band (8.2-12.4 GHz). The dissipation of total EMI SE, EM wave reflection, and EM wave absorption are expressed by $SE_T$, $SE_R$, and $SE_A$. The calculation process is concluded using equations (3)-(8).

$$R + A + T = 1 \quad (3)$$

$$R = |S_{11}|^2 \quad (4)$$

$$T = |S_{12}|^2 \quad (5)$$

$$SE_T = SE_R + SE_A = 10log_{10}(1/T) \quad (6)$$

$$SE_R = 10log_{10}(1/(1-R)) \quad (7)$$

$$SE_A = 10log_{10}((1-R)/T) \quad (8)$$

Where *R*, *A*, and *T* represent the reflection, absorption, and transmission coefficient respectively. $S_{11}$ and $S_{12}$ are measured scattering parameters.

The tensile test of ANF, MXene, ANF/MXene, and ANF/MGN membrane was performed using an electronic universal testing machine (SANS Instruments, CMT 7204). The impact force of free-falling weight was measured by a flat diaphragm type weighting sensor (Bengbu Tongli Sensing Company, CGQ-PM, range 0-200 kg, China) and impact acquisition instrument (Zhongxinleitai, CB-CJS TX-500, China). The open circuit voltage of TENG under cyclic finger pressing was measured using an electrochemical workstation (CH Instruments, CHI850D, China).

## 3. Results and discussion

### 3.1 Characterization of flexible membranes, SSG, and SSG composites

The XRD spectra of ANF, MXene, and ANF/MXene are shown in Fig. 5a. The ANF membrane exhibits the diffraction peaks of (110) and (004) crystal planes at 19.5° and 29.4°. As to the $Ti_3C_2T_x$ MXene membrane, the diffraction peaks of (002) and (004) crystal planes are exhibited at 6.9° and 14.1°. When the two membranes are fabricated as double-layer structure, the diffraction peaks of (002) and (004) of $Ti_3C_2T_x$ shift to 5.9° and 17.4° and the intensity of the peaks declines.

Fig. 5b presents the FT-IR spectra of ANF, MXene, ANF/MXene, and ANF/MGN membrane. The spectrum of ANF exhibits characteristic peaks at 3383 $cm^{-1}$ and 1577 $cm^{-1}$ corresponding to the stretching vibration and deformation of N-H and a characteristic peak at 1682 $cm^{-1}$ corresponding to the stretching vibration of C=O. As to the MXene membrane, the spectrum exhibits peaks at 3745 $cm^{-1}$ and 1428 $cm^{-1}$, indicating the vibration absorption of -OH and C-O. The FT-IR spectrum of ANF/MXene exhibits characteristic peaks of both ANF and MXene, presenting 3308 $cm^{-1}$ and 1372 $cm^{-1}$ corresponding to the stretching vibration and deformation of N-H, and vibration absorption of -OH and C-O at 3701 $cm^{-1}$ and 1592 $cm^{-1}$. Due to the introduction of GO, the spectrum of ANF/MGN membrane exhibits characteristic peaks at 2986 $cm^{-1}$ (C-H), 2899 $cm^{-1}$ (-$CH_3$), 2298 $cm^{-1}$ (C=C), and 1065 $cm^{-1}$ (C-O-C). There are no obvious characteristic Ni peaks due to the absence of chemical reactions with

NiNCs.

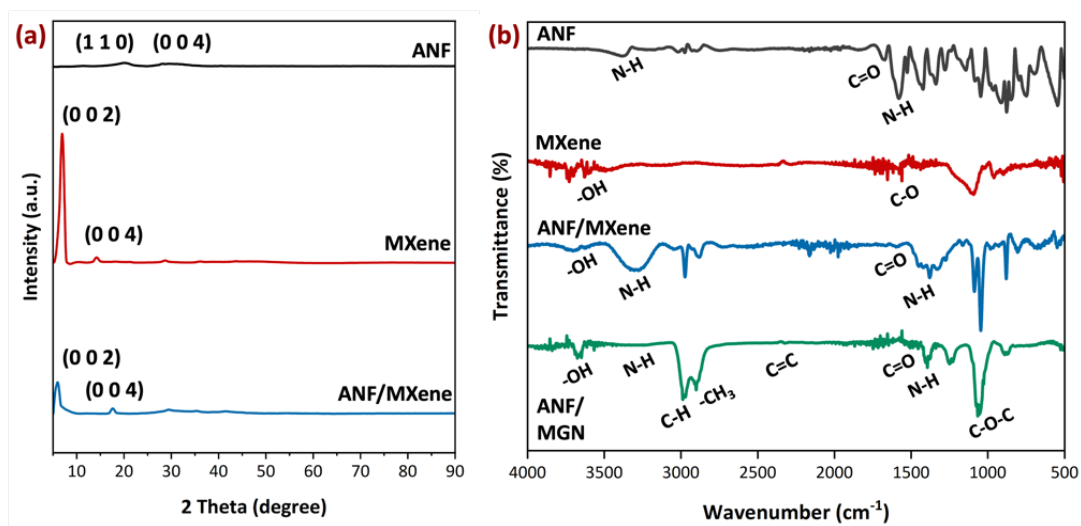

Fig. 5 *(a) XRD spectra of ANF, MXene, and ANF/MXene membrane; (b) FT-IR spectra of ANF, MXene, ANF/MXene, and ANF/MGN membrane*

The wide-scan XPS spectra of ANF, MXene, and ANF/MXene membrane are shown in Fig. 6a. The spectra are revised to the peak of C1s at 284.6 eV. As to the XPS spectrum of ANF, the peaks of C1s, O1s, and N1s are shown clearly due to the element composition of PPTA. Due to the existence of C and Ti elements in $Ti_3C_2T_x$, peaks of C1s, Ti2s, Ti2p, Ti3s, and Ti3p are shown clearly. Additionally, peaks of F1s and O1s can be observed in the spectrum. That is because the etching process introduces surface functional groups such as -OH, -F, etc., leading to corresponding peaks in the spectrum. After the combination of ANF/MXene double-layer, peaks of both ANF and MXene single membrane can be observed in the spectrum. Peaks of O1s and C1s exhibit an obvious rise compared with spectra before the double-layer combination. Narrow spectra of high resolution are shown in Fig. 6b and c. As to C1s spectra of ANF and MXene/ANF membranes, peaks of C-C and C-N do not exhibit obvious shift, with the binding energy of 284.8 eV and 285.5 eV in ANF spectrum, and 284.7 eV and 285.7 eV in MXene/ANF spectrum. However, after the double-layer combination, the peak of C=O shifts from 288.1 eV to 288.9 eV. Additionally, peaks of C-O (287.4 eV) and C-Ti (282.5 eV) appear as the component of C1s peak. The appearance of C-O indicates the higher hydrogen bonding reaction probability between the MXene and ANF layers. In the spectrum of O1s for the MXene membrane, the peak of C-Ti-$O_x$ at 530.4 eV and C-Ti-OH at 531.9 eV can be observed. While in the spectrum of O1s for MXene/ANF membrane, the peak of C-Ti-$O_x$ at 530.8 eV and C-Ti-OH at 531.7 eV can be observed, exhibiting little shift after the double-layer combination. As to peaks of O-Al and O-Ti, shift can be obtained from 532.9 eV to 532.3 and 529.6 eV to 528.0 eV, respectively. The peak at 534.0 eV in the MXene O1s spectrum and the peak at 534.5 eV in the MXene/ANF O1s spectrum indicate the absorbed water in the membrane. In the MXene/ANF spectrum, an O=C peak at 532.8 eV appears, indicating higher hydrogen bonding reaction probability between the MXene and ANF layer.

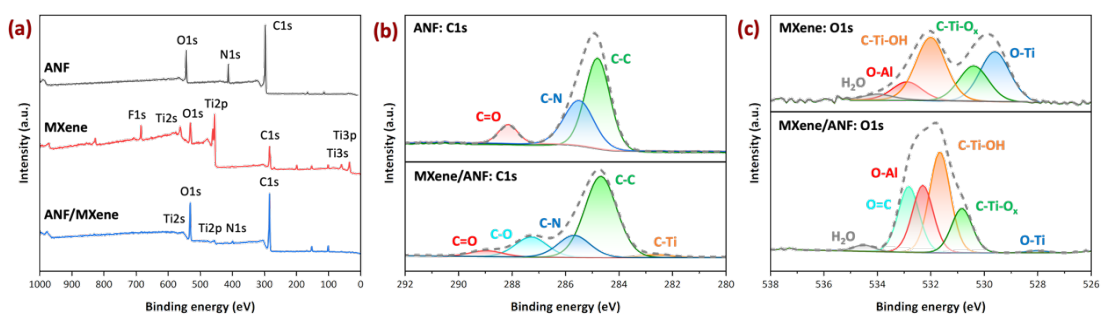

Fig. 6 *(a) XPS wide scan spectra of ANF, MXene, and ANF/MXene membrane; (b) XPS narrow scan s C1s peaks for ANF and MXene/ANF membrane, MXene is noted first to indicate that MXene layer is the upper layer when taking on the characterization; (c) XPS narrow scan O1s peaks for MXene and MXene/ANF membrane*

The PU oriented foam fabricated in this work exhibits obvious short-range order porosity (Fig. 7a). SSG exhibits a transparent and light-yellow appearance (Fig. 7b). After the introduction of SSG into the PU matrix, the SSG/PU composite exhibited lustre and improved elasticity (Fig. 7c). The elasticity of PU foam, SSG/PU composite, and SSG was tested by bending the sample and releasing. The pure PU exhibited moderate elasticity (Fig. 7d-g). The deformation was restored by a half after releasing for 5 seconds. After that, the restoration slowed dramatically. When the sample was released for 40 seconds, the restoration almost stagnated, remaining at a quarter of the deformation. Therefore, the resilience is inadequate. After being released from bending, SSG exhibit plasticity without any restoration (Fig. 7h-k). The cold flow effect can be observed after 40 seconds. As to SSG/PU composite, the elasticity improved compared with pure PU foam (Fig. 7l-o). The deformation was restored as fast as SSG foam for the first 5 seconds. Different from PU foam, the restoration was almost finished when the sample was released for 40 seconds, indicating the modification of insufficient resilience of PU foam. To test the modification of the cold flow effect, SSG and SSG/PU composite under the same dimension were placed on the edge of the culture dish. After 6 minutes, SSG exhibited an obvious cold flow effect while SSG/PU composite remained the same, indicating the dimensional stability of SSG/PU composite.

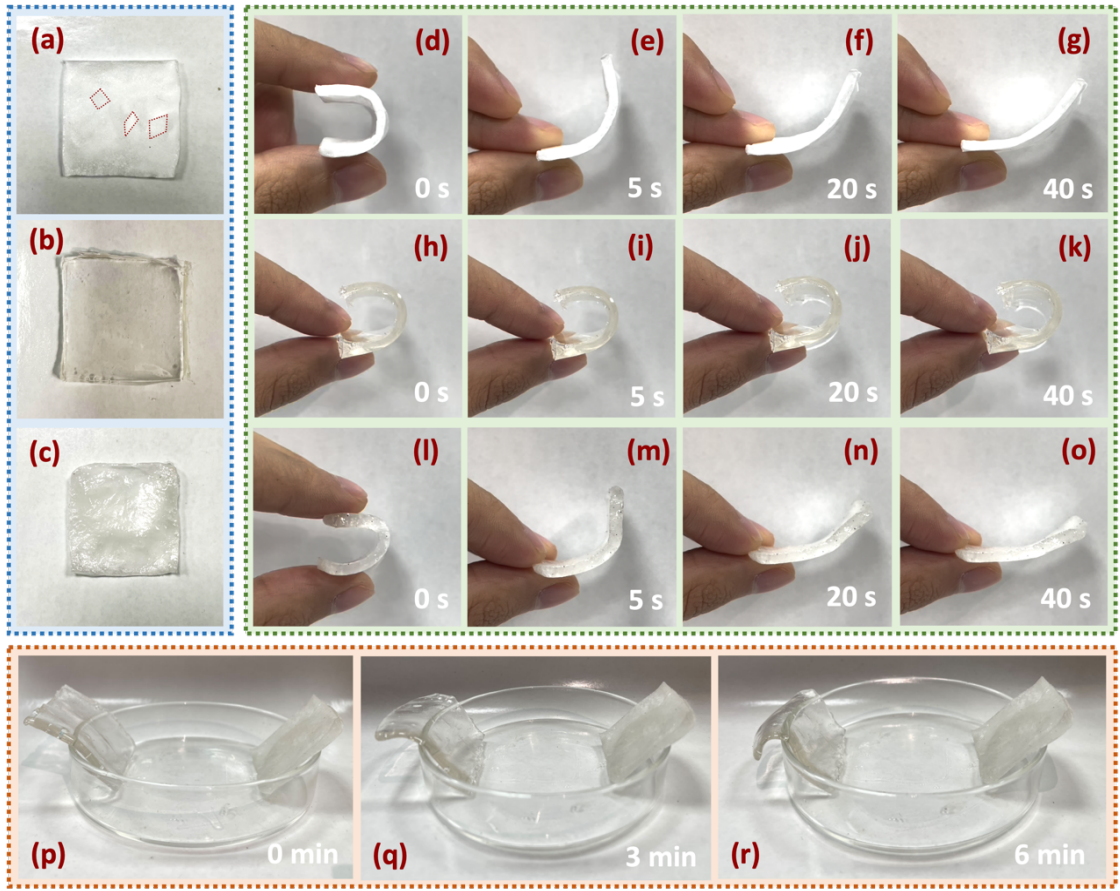

Fig. 7 *The appearance of (a) PU oriented foam, red region indicates certain short-range order porosity, (b) SSG, and (c) SSG/PU composite; the bending-releasing test of (d-g) PU oriented foam, (h-k) SSG, and (l-o) SSG/PU composite; (p-r) the cold flow effect test of SSG and SSG/PU composite*

The rheology diagrams of SSG and SSG/PU composite are plotted after characterization. As to SSG, when the angular frequency increases from 0.1 to 10 rad/s, the storage modulus increases and the loss modulus increases first and then decreases, indicating the feature of the shear stiffening effect (Fig. 8a) [44, 45]. After the introduction to the PU matrix, the SSG/PU composite remains the shear stiffening characteristic (Fig. 8b). Compared with pure SSG, the shear stiffening effect exhibits a certain decline. The stiffening region remains 0.1 to 10 rad/s. Higher initial modulus exhibits improved modulus of SSG/PU composite compared with SSG. Additionally, lower modulus changing amplitude also indicates improved structure stability compared with pure SSG. Therefore, this SSG/PU composite can be applied as an energy absorbing core in the flexible sandwich structure.

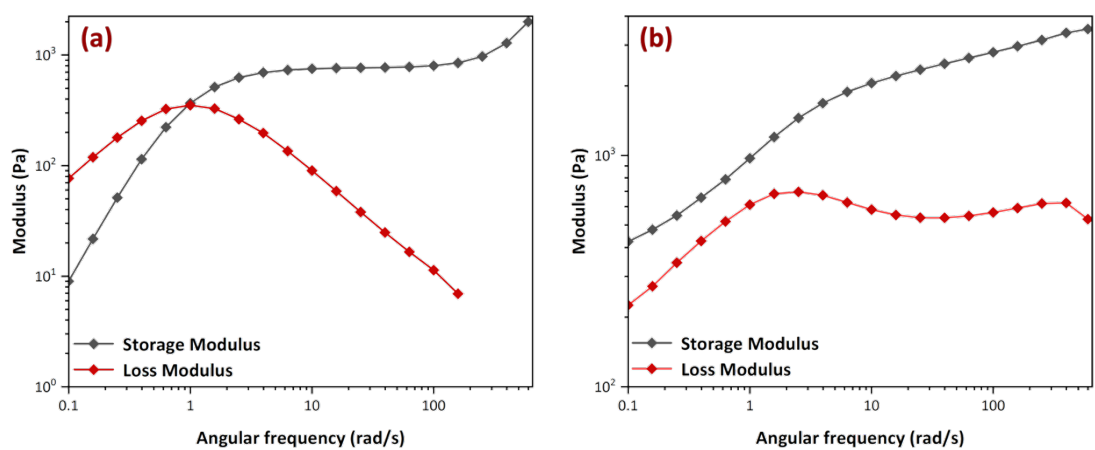

Fig. 8 *Rheology diagram of pure (a) SSG and (b) SSG/PU composite*

**3.2 Morphologies of PU matrix, SSG/PU composites, and flexible membranes**

The section parallel to the oriented holes of PU foam and SSG/PU composite was observed under SEM. As shown in Fig. 9a, the oriented porous structure appears explicitly, indicating the occupancy of ice before vacuum freeze drying. Due to the existence of the microporous structure, SSG can fill in the PU matrix thoroughly by vacuum filtration, leading to SSG/PU composite, shown in Fig. 9b. Additionally, the oriented structure of the matrix still can be observed after the filling with SSG.

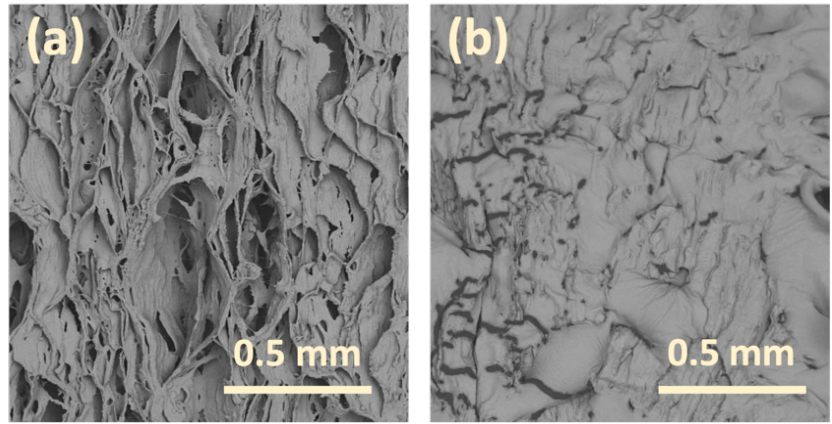

Fig. 9 *SEM figures of sections of (a) PU matrix and (b) SSG/PU composite after 40 times magnification (marker bars = 0.5 mm)*

The ANF dispersion and aqueous MXene used in this work exhibit different colours under different concentrations. After stirring for 7 days, the ANF solution exhibits a transparent dark red appearance. In a 5 mL sample vial, the colour changes to orange (Fig. 10a). After being diluted by DMSO and deionized water in the volume ratio of 1:9:1, the ANF dispersion has in light yellow appearance (Fig. 10b). As to the aqueous MXene used in this work, it exhibits a black appearance, indicating a high concentration of $Ti_3C_2T_x$ (2.99 mg/mL). However, after being diluted for 10 times and 100 times, the green colour [35] of $Ti_3C_2T_x$ can be observed (Fig. 10c, d). The liquid in 5 mL sample vial exhibits light beam scattering under laser irradiation, indicating a certain size range of the dispersoid.

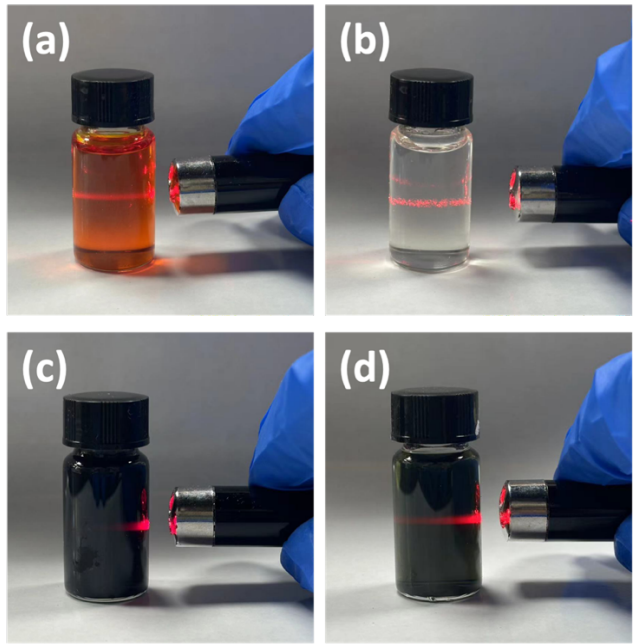

Fig. 10 *Light beam scattering under laser irradiation of (a, b) ANF dispersion and (c, d) aqueous MXene in 5 mL sample vials*

The membranes fabricated in this work were cut into rectangular shapes before application. As pure single-layer membranes (MXene and ANF membranes), the two sides do not exhibit obvious differences. As shown in Fig. 11a, some rough patterns can be observed on the surface, which are attributed to the rough surface of the vacuum filtration device. However, these rough patterns do not appear on the surface of the ANF membrane (Fig. 11b). After the assembly of double-layer membranes, different morphology can be observed at different sides. From the conductive side of ANF/MXene and ANF/MGN membranes (Fig. 11c, e), rough patterns can be observed. However, the ANF side of these two membranes still exhibits smooth surface morphology (Fig. 11d, f).

After magnifying 200 times under SEM, wrinkle patterns can be observed on the surface of the conductive side (Fig. 11a, c, e). From pure MXene (Fig. 11a) to

ANF/MXene (Fig. 11c), and to ANF/MGN (Fig. 11e), the wrinkle patterns on the conductive layer gradually disappear. After magnifying 2500 times under SEM, ignoring impurities on the surface, both the conductive side and ANF side exhibit smooth appearances. Among all the surfaces presented, the surface of MGN exhibits the most obvious micro-roughness (Fig. 11e). Specially, NiNCs can also be observed on the MGN surface.

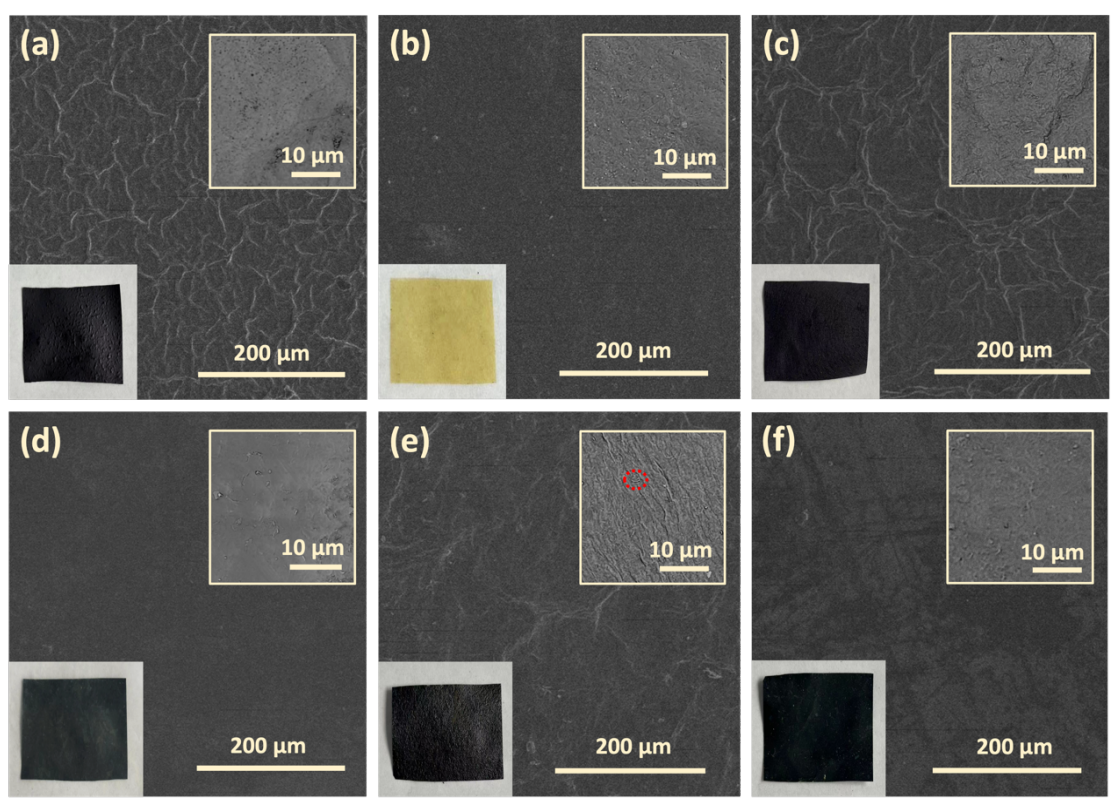

Fig. 11 *The appearance of macroscale, 200 times magnification SEM (marker bars = 200μm), and 2500 times magnification SEM (marker bars = 10μm) of (a) MXene membrane, (b) ANF membrane, (c) ANF/MXene membrane conductive side, (d) ANF/MXene membrane ANF side, (e) ANF/MGN membrane conductive side, red circle indicates the NiNC, (f) and ANF/MGN membrane ANF side*

The sections of ANF, MXene, ANF/MXene, and ANF/MGN membranes under

SEM are shown in Fig. 12a-d. As it is shown clearly, the nano-layer structure of the membranes is confirmed. Shown in Fig.12c, upper layer is the MXene layer, and lower layer is the ANF layer. Shown in Fig. 12d, upper layer is the MGN layer, and lower layer is the ANF layer. Due to the existence of GO, the thickness of ANF/MGN is much higher than other flexible membranes. Additionally, NiNCs can be observed within MGN layer, linking different nanolayers. After the introduction of GO and NiNCs, the conductive layer becomes more brittle, leading to delamination within MGN layer. Sections of A/S/A, A/M/S/M/A, and A/MGN/S/MGN/A structures under SEM are shown in Fig. 12e-g, indicating obvious sandwich structure appearance.

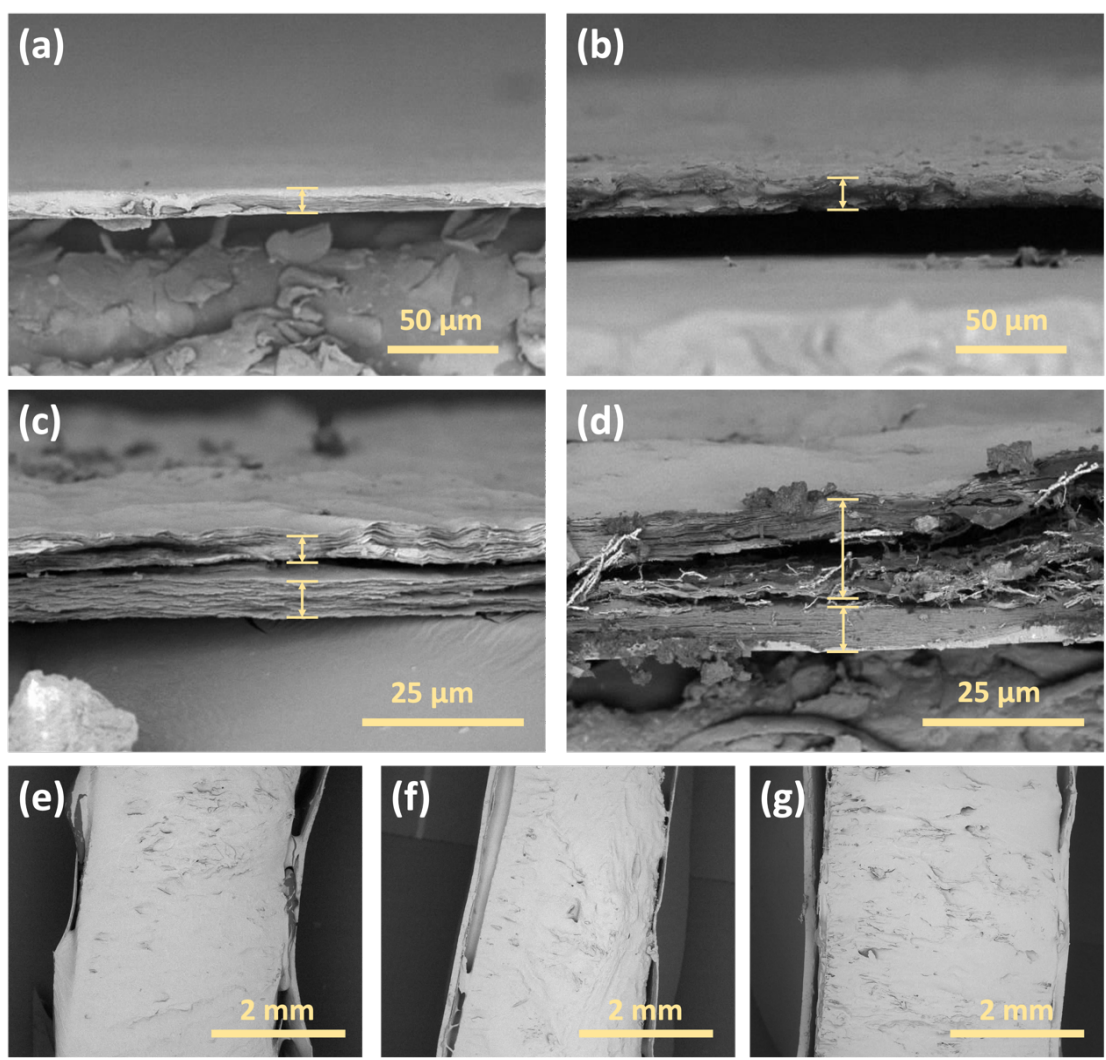

Fig. 12 *The section of (a) ANF and (b) MXene (marker bars = 50 μm), (c) ANF/MXene and (d) ANF/MGN membranes (marker bars = 25 μm); sections of (e) A/S/A, A/M/S/M/A, and A/MGN/S/MGN/A structures (marker bars = 2 mm)*

**3.3 Electrical conductivity of flexible membranes and electromagnetic interference shielding characterization of flexible sandwich structures**

As conductive layers in the flexible sandwich structure, the electrical conductivity of the MXene membrane, ANF/MXene membrane, and ANF/MGN membrane were measured. All the measurement was taken on the conductive side of the membranes (avoid insulated ANF layer). Each membrane was measured 6 times. To determine whether the mean value difference of each data series is statistically significant, Student's t-test [61] was applied for each data series pair (Table 2). Compared with the chosen predetermined significance level of 0.05, all the p-values exhibit a degree of $10^{-7}$, which is lower than 0.05. Therefore, the null hypothesis can be rejected, and the mean value of each sample differs significantly from each other.

Table 2. *p-value of data series pair of Student's t-test*

| Series | MXene *vs* A/M | A/M *vs* A/MGN | MXene *vs* A/MGN |
| --- | --- | --- | --- |
| p-value | $1.008 \times 10^{-7}$ | $1.306 \times 10^{-7}$ | $1.306 \times 10^{-7}$ |

The conductivity of MXene, ANF/MXene, and ANF/MGN exhibit the regulation of decline (Fig. 13), all the data was reserved to 4 significant figures. The mean value conductivity of the MXene membrane is 6464 S/m, exhibiting excellent metallic conductivity of $Ti_3C_2T_x$. As to the ANF/MXene membrane, this value decreases to 368.4 S/m, because of the introduction of the insulated ANF layer. After the introduction of GO and NiNCs, the conductivity decreases dramatically to 0.001323 S/m. The first reason is the reduction in the amount of MXene (40 mg to 35 mg), a second reason is that the conductivity of GO is substantially lower than graphene. The oxygen-containing functionalities destroy the major π bond among the graphene surface, lowering the conductivity [62].

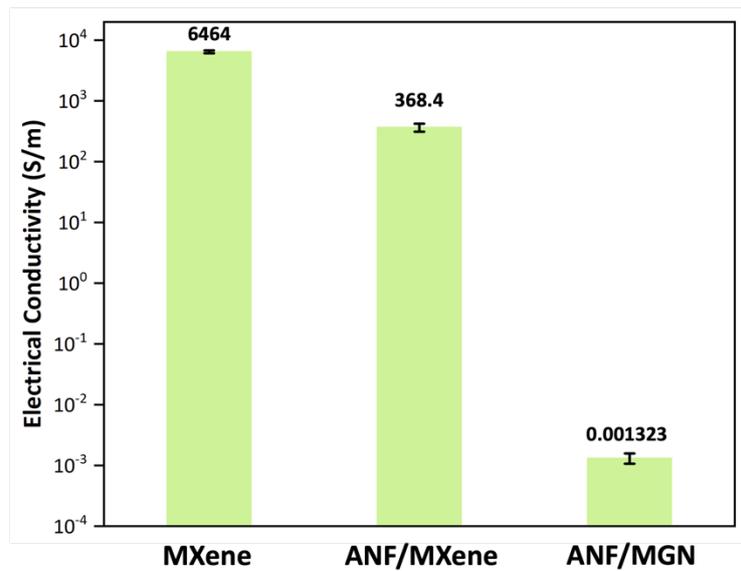

Fig. 13 *Mean electrical conductivity of MXene, ANF/MXene, and ANF/MGN membrane*

As for polymer EMI shielding composites, the main shielding mechanisms of total shielding effectiveness ($SE_T$) are reflection ($SE_R$) and absorption ($SE_A$) [63]. The shielding performance of SSG, A/S/A, A/M/S/M/A, and A/MGN/S/MGN/A structures were characterized. Especially, the structure of S/M/A (incident waves come to the S layer first) and A/M/S (incident waves come to the A layer first) were also characterized. $SE_R$, $SE_A$, and $SE_T$ at X-band (8.2-12.4 GHz) are shown in Fig. 14, and all the data was reserved in three significant figures.

Due to the insulation of SSG and ANF, both pure SSG layer and A/S/A layer exhibit minimal EMI SE (~ 3 dB). EMI SE of A/M/S/M/A exhibits 62.6 dB with a thickness of

4.19 mm (0.04 mm for a single ANF/MXene layer). A/MGN/S/MGN/A exhibits a slight decrease compared with the A/M/S/M/A structure. The EMI SE is 59.9 dB with a thickness of 4.48 mm (0.11 mm for a single MGN layer). Comparing double conductive layer structure (DCLS) with single conductive layer structure (SCLS), SCLS exhibits an obvious shielding effectiveness decline of about 50%, indicating the advantage of double conductive layer. As to SCLS with EM waves from different sides, there is no obvious difference with $SE_T$ (28.3 dB and 30.2 dB). However, the ratio of $SE_T$ is different. For S/M/A, $SE_R$ occupies only 2.5% in $SE_T$, while the $SE_R$ of A/M/S occupies 44.3 % in $SE_T$.

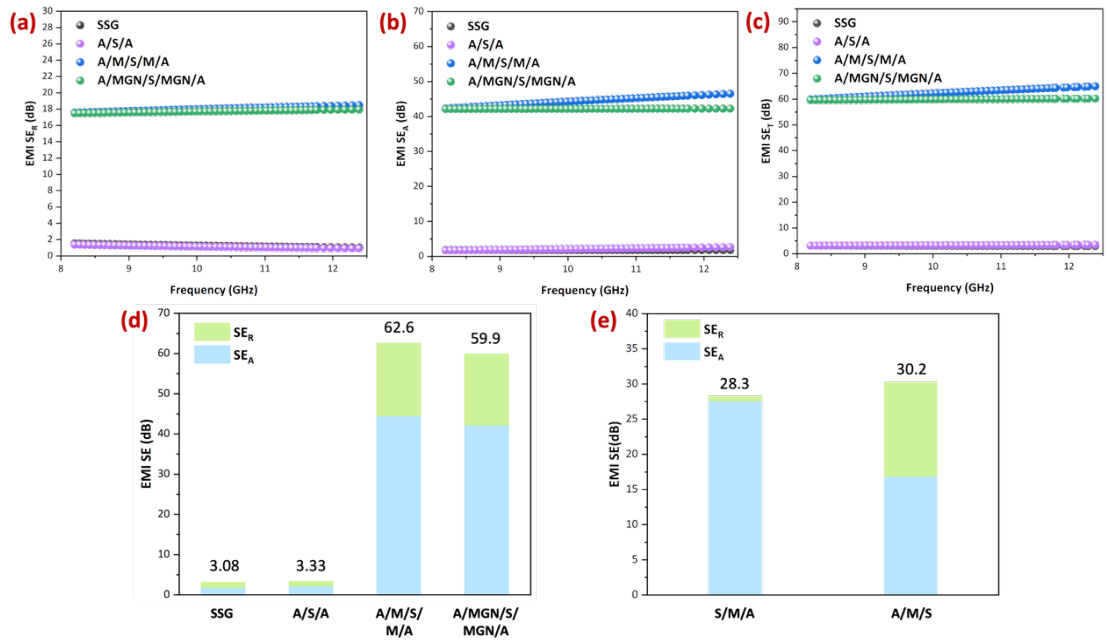

Fig. 14 *(a) EMI $SE_R$, (b) EMI $SE_A$, (c) EMI $SE_T$ at X-band; the mean value of EMI $SE_T$ with $SE_R$-$SE_A$ composition of (d) DCLS and (e) SCLS*

The EMI shielding mechanism of the flexible sandwich structure can be disassembled into three parts (Fig. 15):

Firstly, when incident EM waves come to layer materials, reflection happens due to impedance mismatch [64]. The surface reflection of the top layer is $SE_R$, which is relatively weak and only occupies a low percent of $SE_T$. After the transmission of the top layer, internal interface multireflection happens to attenuate EM waves, which can be concluded as part of $SE_A$. Additionally, due to a huge impedance mismatch compared with the conductive layer and other layers, the reflection on the conductive layer surface is the most significant.

Secondly, the conductive layer also plays other vital roles. Both MXene and MGN exhibit nano-layer structures. Therefore, when incident EM waves come to the conductive layer, repeated reflection between nanolayers happens [36]. This microscopic multireflection contributes to $SE_A$. Simultaneously, EM waves interact with high-density electrons in the MXene lattice to induce a current, which can be concluded as ohmic loss for absorbing attenuation [36]. As for NiNCs in the layer, eddy currents [65] will be generated under a magnetic field, attenuating the energy of EM waves. However, though it has the highest thickness, A/MGN/SSG/MGN/A exhibits a slight decrease in EMI SE compared with A/M/S/M/A. That is because the EMI SE performance of the GO/NiNCs system used in this work is lower than MXene, another reason is insufficient NiNCs (only 20 mg).

Thirdly, when passing through the SSG/PU composite layer, EM waves transmit from the micro-scale holes. During this process, reflection happens when EM waves reach the edge of micro-holes, however, this attenuation is weak. Therefore, SSG and A/S/A only exhibit EMI SE of ~3 dB.

As to SCLS of S/M/A, surface reflection is lower than that of A/M/S due to lower conductivity of SSG/PU layer compared with ANF/MXene layer. Incident EM waves reflect on the edge of micro-holes when transmitting the SSG/PU layer. The waves reflect on the MXene layer and reflected EM waves encounter the reflection attenuation of micro-holes of the PU matrix again, leading to more absorption percent. As to SCLS of A/M/S, relatively higher percentage of $SE_R$ can be observed due to higher conductivity of ANF/MXene layer. Transmitted EM waves reflected within nano-layer structures and the edge of micro-holes of SSG/PU layer. The amount of $SE_R$ of A/M/S structure is similar to that of A/M/S/M/A structure.

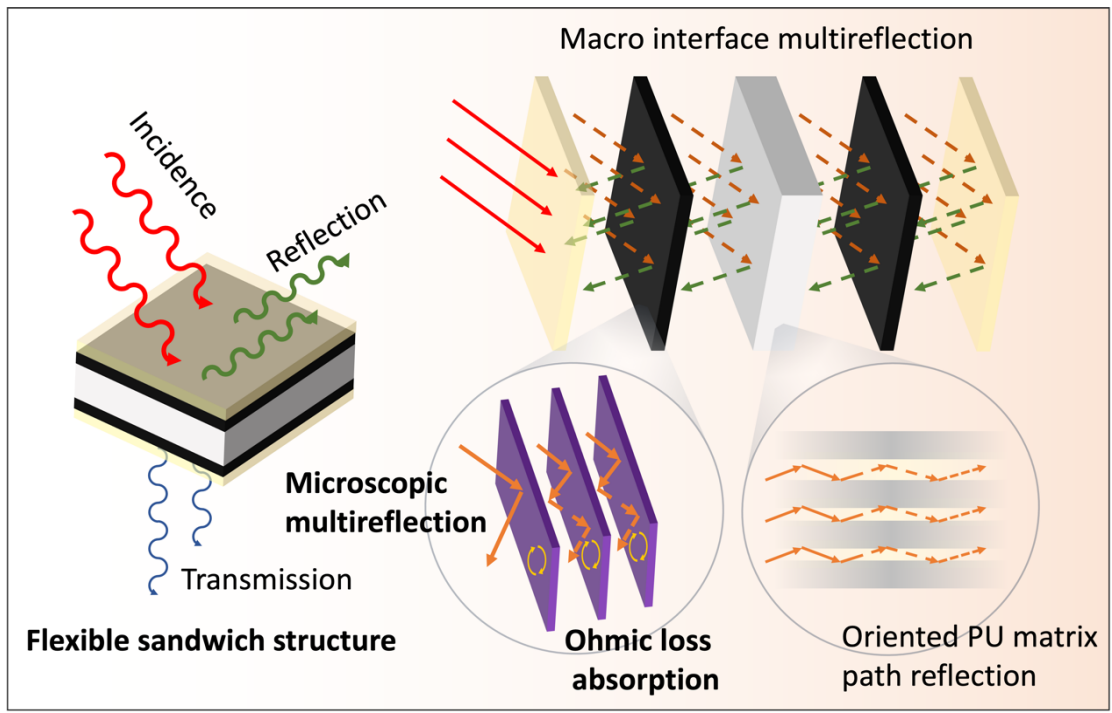

Fig. 15 *Schematic illustration of EMI shielding mechanism of flexible sandwich structure*

## 3.4 Tensile testing of flexible membranes and impact resistance characterization of flexible sandwich structures

One of stress-strain curves of ANF, MXene, ANF/MXene, and ANF/MGN

membranes are presented in Fig. 16b-e. The ANF membrane exhibits a tensile strength of 164.6 MPa and fracture strain of over 9 % (Fig. 16b). The vacuum-filtrated ANF membrane can be seen as ANF aerogel [66]. In the aerogel network, ANFs interact with each other by hydrogen bonding, and among the nanofiber, the PPTA chains interact with each other by hydrogen bonding, resulting in high strength and toughness of both the nanofiber and the membrane [40, 42, 66]. However, the MXene membrane exhibits a brittle characteristic, resulting a tensile strength of only 13.93 MPa and a fracture strain of below 4.5 % (Fig. 16c), this is because most of the interaction between $Ti_3C_2T_x$ nanolayers exist in interlamination. When the stress is applied in the radial direction of $Ti_3C_2T_x$ layers, they can easily break apart. After the assembly of the ANF/MXene double-layer membrane, the mechanical properties are improved compared with pure MXene. During double-layer assembly, the ANF layer attaches to the MXene layer by hydrogen bonding. The tensile strength of ANF/MXene membrane achieves 57.15 MPa, exhibiting over 3 times increase (Fig. 16d). The combination of ANF membrane also enhances the fracture strain of MXene, leading to over 6 % fracture strain of ANF/MXene and corresponding toughness modification. The stress-strain curve of ANF/MXene exhibits a disturbance at 4.6 % strain, indicating two steps during the tensile test of the ANF/MXene membrane. After the strain reaches 4.6 %, the MXene exhibits cracks first. Due to the moderate testing speed of the device (10 mm/min), the whole membrane fails at 6.3 % strain finally. Therefore, two fracture cracks can be observed in a single ANF/MXene membrane sample (Fig. 16f). The same fracture mechanism can also be observed at ANF/MGN tensile testing. However, the disturbance appears early at 0.7 %. The first reason is that the introduction of NiNCs hinders the toughness of the flexible layer [28]. Apart from that, the interaction between GO nanolayers and MXene nanolayers, and the interaction between the conductive layer and ANF layer are lower. The ANF/MGN membrane only achieves 26.87 MPa tensile strength and 2.5% final tensile strain (Fig. 16e), indicating lower toughness and strength compared with ANF/MXene.

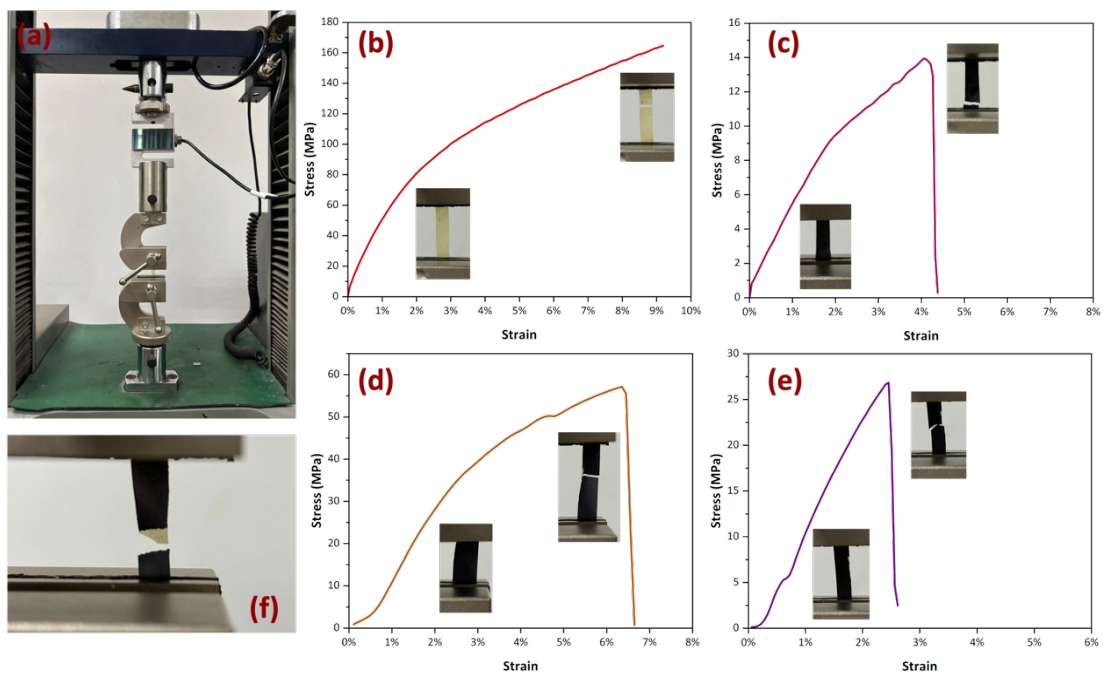

Fig. 16 *(a) The electronic universal testing machine; one of stress-strain curves of (b) ANF, (c) MXene, (d) ANF/MXene, and (d) ANF/MGN membranes; (f) two cracks seen in the ANF/MXene membrane*

Also, the independent two-sample Student's t-test was applied for MXene vs ANF/MXene and ANF/MXene vs ANF/MGN data series for both tensile stress and strain. As shown in Table 3 and Table 4, all the p-values are lower than the chosen predetermined significance level of 0.05. Therefore, the null hypothesis can be rejected, and the mean value of each sample differs significantly from each other, indicating efficient strength modification of the ANF layer in ANF/MXene and strength decline of ANF/MGN. Mean values and standard deviation of tensile stress and tensile strain are shown in Fig. 17. The fracture strength and fracture strain are improved by the ANF layer. And ANF/MGN exhibits brittle properties compared with ANF/MXene.

Table 3. *p-value of tensile stress data series pair of Student's t-test*

| Series | MXene *vs* ANF/MXene | ANF/MXene *vs* ANF/MGN |
| --- | --- | --- |
| p-value | $2.400 \times 10^{-4}$ | $1.177 \times 10^{-3}$ |

Table 4. *p-value of tensile strain data series pair of Student's t-test*

| Series | MXene *vs* ANF/MXene | ANF/MXene *vs* ANF/MGN |
| --- | --- | --- |
| p-value | 0.04982 | 0.01454 |

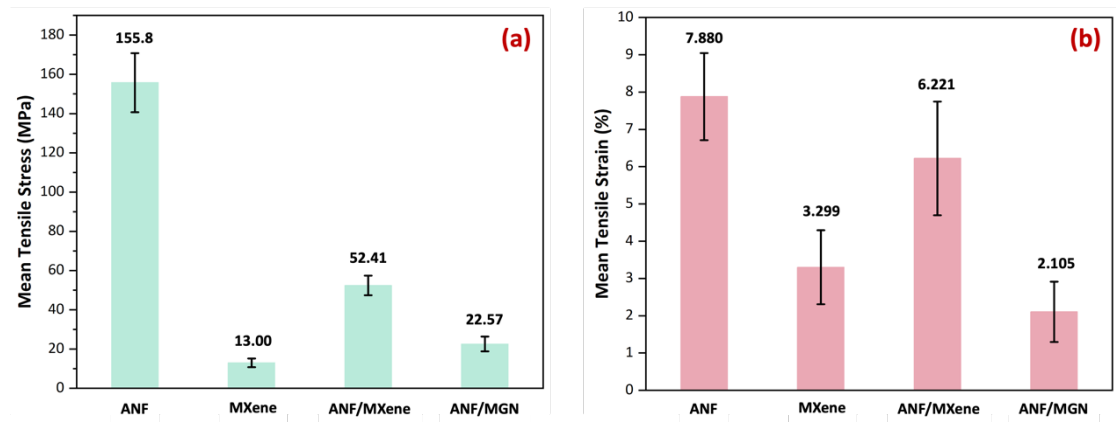

Fig. 17 *(a) Mean values of (a) membrane tensile stress and (b) membrane tensile strain*

The impact test used constructed equipment (Fig. 18a) for ball-falling experiment. The red rectangle (1) is the control button, in which the red one is the signal light, the yellow one is the button to switch on the electromagnet, and the blue one is the button to switch off the electromagnet. Rectangle (2) is the electromagnet and attached steel impact ball. The diameter of the impact ball is 49.72 mm, and the weight is 500.03 g. Rectangle (3) is the flat diaphragm-type weighting sensor. The sample was placed on the top of the sensor. Rectangle (4) is the impact acquisition instrument. Rectangle (5) is the computer running the operation (the operation was supplied by the manufacturer). During the test, the impact ball was released after the electromagnet was switched off. The sample was impacted, and the force was measured by the sensor. The signal was collected by the acquisition instrument and sent to the computer, giving the value of impact force.

The impact forces to different samples of free-falling steel ball with ball-falling heights increasing from 20 to 50 cm were collected (Fig. 18b, c, d). Samples containing SSG core exhibited obvious impact force resistance. As the impact force increases from 28 to 69 N, the force can be attenuated to below 20 N, the attenuation increases with increasing impact force (Fig. 19). As to A/S/A, the attenuation percent increases from 53.6 % to 72.5 % (Fig. 19b). A/S/A exhibits better performance compared with pure SSG, indicating the impact resistance of the ANF membrane. However, as to A/M/S/M/A and A/MGN/S/MGN/A, the degree of attenuation exhibits slight declines because of the stiffness of MXene.

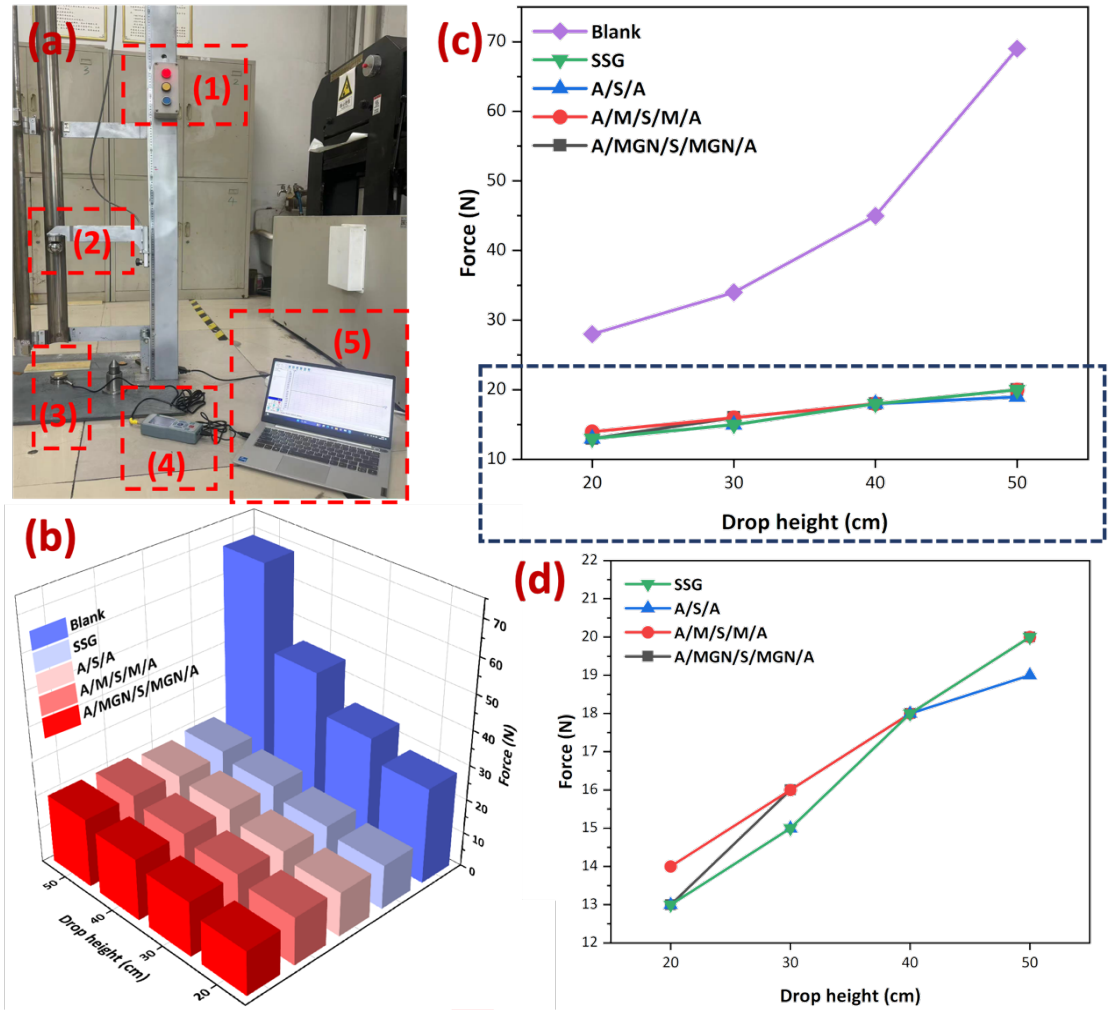

Fig. 18 *(a) Impact test equipment constructed; (b, c, d) impact forces to different samples under ball-falling height increasing from 20 to 50 cm*

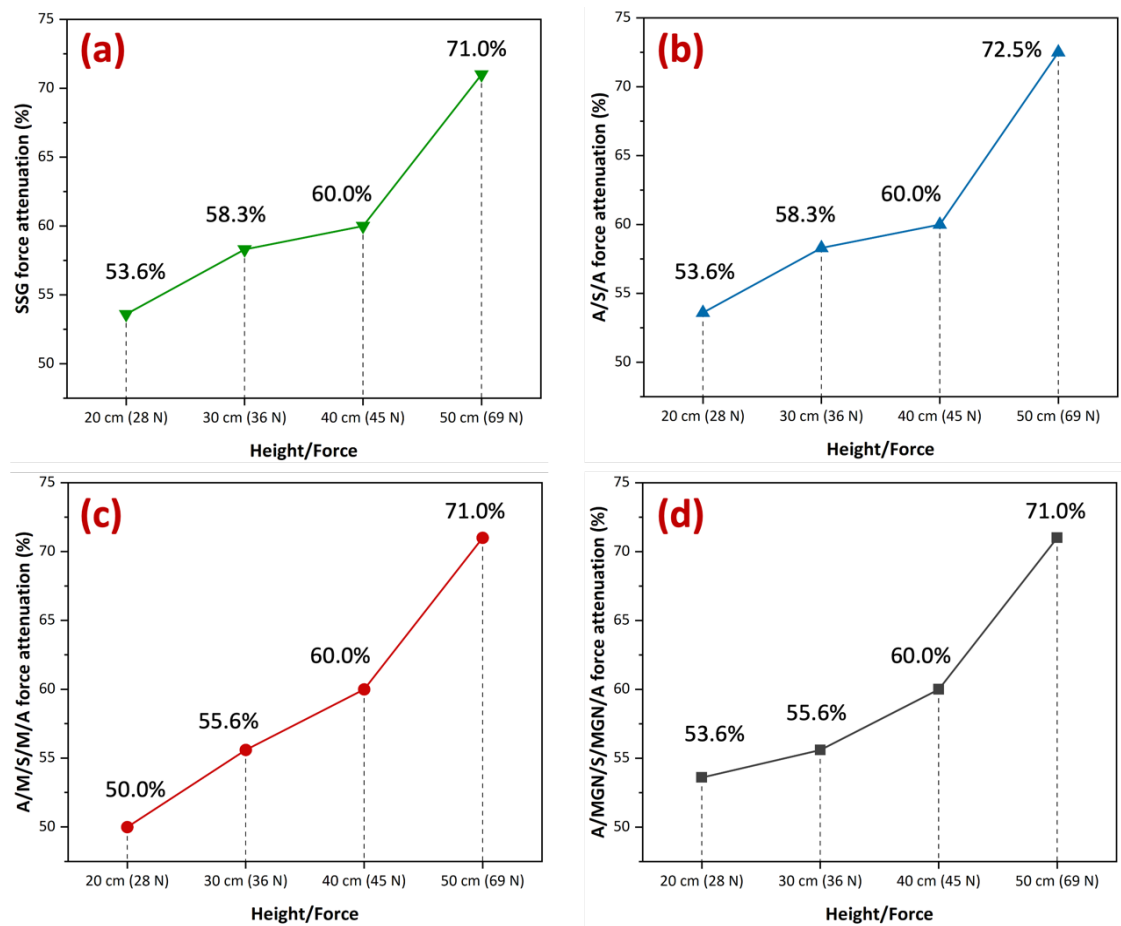

Fig. 19 *Impact force attenuation of (a) pure SSG, (b) A/S/A, (c) A/M/S/M/A, and (d) A/MGN/S/MGN/S/A with the increase of ball falling height or impact force*

The mechanism of impact resistance is shown in Fig. 20. The intermolecular crosslinking of SSG is -B-O- bonding, which is integrated by the p orbital of the B atom and electron of the O atom, indicating dynamic bond characteristics [44, 48]. Therefore, different results can be observed under stimuli of different strain rates. Under low strain rate, the dynamic -B-O- bonding separates apart, exhibiting plastic deformation of SSG. However, when subjected to high strain rate, the dynamic bonding integrates to form a high-density network, blocking the movement of polymer chains. During high strain rate impact, the polymer gel deforms to attenuate the incident energy. Simultaneously, some of the gel forms a powder, leaving white trails on the gel, which is called whitening. This phenomenon also attenuates impact energy [48]. Additionally, the oriented PU matrix contributes to impact force resistance. The -B-O- dynamic bonding returns when the high strain rate disappears. As to A/S/A structure, apart from SSG, the ANF network membrane also attenuates impact energy. When encountering to high strain rate, ANF networks exhibit elasticity to generate conical deformation [43]. Cracks are also generated if the stress is beyond the strength of ANF networks, attenuating impact energy. After introducing the conductive layer (MXene or MGN), delamination also contributes to energy attenuation, which happens at the ANF/MXene interface or MGN interior. To conclude, SSG plays the most important role in energy attenuation.

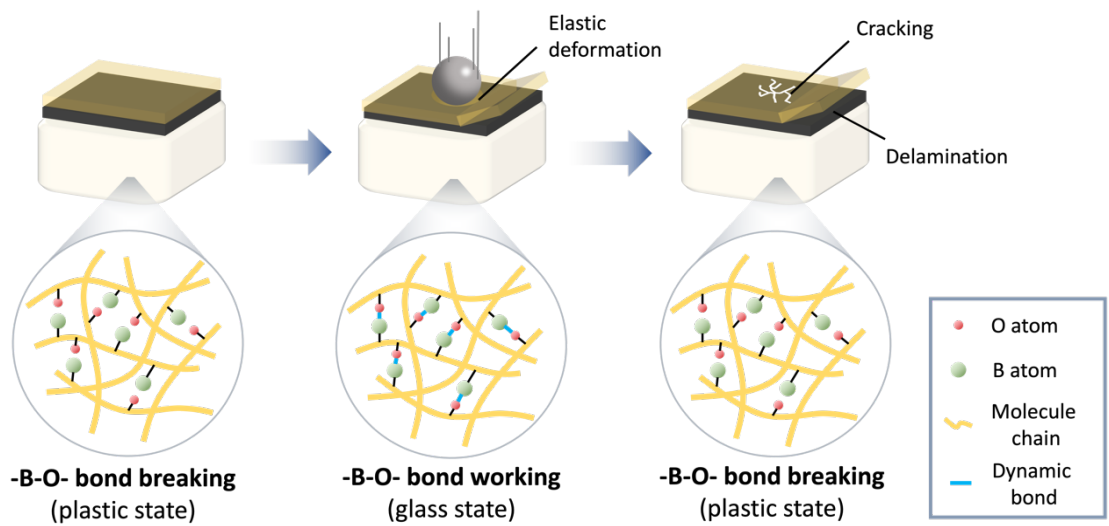

Fig. 20 *Schematic illustration of the mechanism of impact force attenuation of ANF/MXene/SSG sandwich structure.*

The sample morphology before and after 50 cm-ball-falling impact force on SSG (Fig. 21a, b), A/S/A (Fig. 21c, d), A/M/S/M/A (Fig. 21e, f), and A/MGN/S/MGN/A (Fig. 21g, h) was photographed. The impact result of whitening (green rectangle), membrane crack (orange rectangle), and membrane delamination (red rectangle) can be observed.

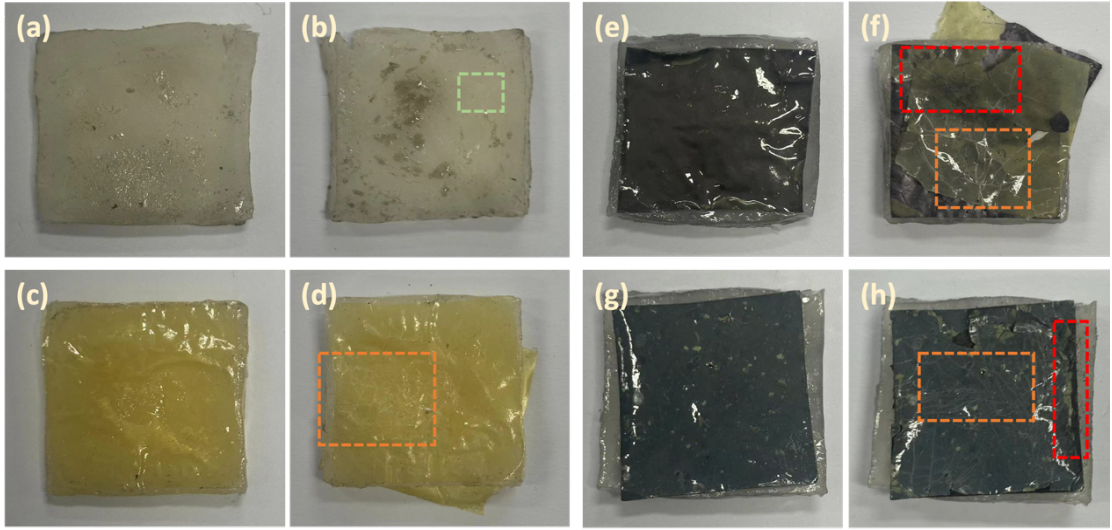

Fig. 21 *Sample morphologies before and after 50 cm-ball-falling impact force of (a, b) SSG, (c, d) A/S/A, (e, f) A/M/S/M/A, and (g, h) A/MGN/S/MGN/A*

**3.5 MGN/PDMS TENG design theory and performance**

The triboelectric nanogenerator was invented according to the triboelectric effect and electrostatic induction [56, 57]. When two tribo-pairs contact with each other, electron transfer happens, leading to charges of opposite signs on two dielectric surfaces. When separating the two tribo-pairs, charges remain on the dielectric surfaces. For multilayer structures, the opposite charges are induced on layers attaching with each single tribo-pair according to electrostatic induction. Therefore, an external current will be generated after connecting two layers with different potentials to balance the potential difference [58]. In this model, the output current is essentially Maxwell's displacement current. The Maxwell's displacement current can be defined as:

$$J_D = \frac{\partial D}{\partial t} = \varepsilon_0 \frac{\partial E}{\partial t} + \frac{\partial P}{\partial t} \quad (3\text{-}1)$$

Where D represents displacement field, $\varepsilon_0$ represents permittivity in vacuum, *E* represents the electric field, and P represents the polarization field.

As to the MGN/PDMS TENG assembled in this work, MXene acts as an electron donor due to its metallic conductivity. The PDMS rubber is an insulator, which accepts electrons to maintain negative charges [67]. To create more surface area, which increases triboelectric charges, NiNCs were introduced into MXene. The existence of NiNCs increases the surface roughness of the MXene membrane. Apart from that, certain researchers discovered that graphene can be applied to exhibit better dispersion of inorganic nanofibers in the system [68]. Considering that, a certain amount of GO was also introduced into the system. Finally, the conductive layer (briefly called the MGN layer) consists of MXene, NiNCs, and GO (the parameter of the MGN layer is already listed before). Therefore, the MGN layer acts as the cathode and PDMS rubber acts as the anode for TENG for energy generation. When the surface of TENG is pressed, the two tribo-pairs contact with each other, generating positive charges on the conductive layer and negative charges on PDMS rubber. After the external force is removed, the two tribo-pairs separate from each other, and the opposite charges remain on the electrodes. Due to electrostatic induction, the ANF layer generates negative charges and copper foil generates positive charges. Simultaneously, electrons transfer from PDMS rubber to copper foil with external wire, exhibiting output current. Therefore, cyclic external pressing force generates a cyclic electric output signal (Fig. 22).

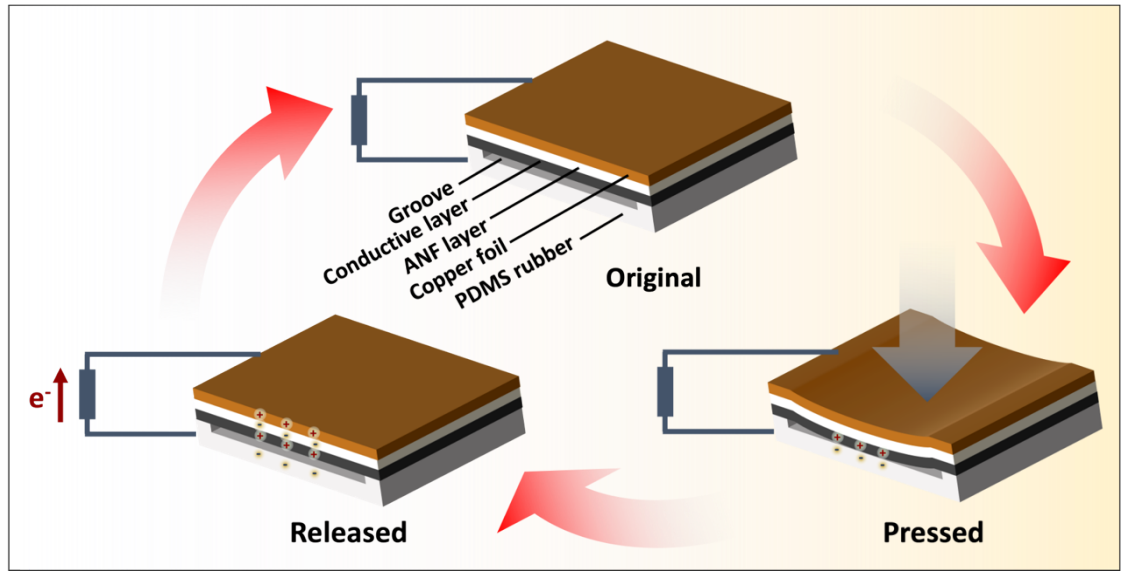

Fig. 22 *Schematic illustration of working principle of MGN/PDMS TENG*

As shown in Fig. 23, the voltage generated by MGN/PDMS TENG under cyclic pressing was measured. The MGN/PDMS TENG was cyclically pressed for 70 s (83 cycles). As shown in Fig. 23a, this MGN/PDMS TENG exhibited relatively stable voltage generation. Cycles during 9-12.4 s and 55-58.5 s were picked out, shown in Fig. 23b and Fig. 23c. Apart from multiple cycles, voltages generated under different contacting areas were also measured. When the surface was tapped by a single finger, about 0.4 V voltage was generated for every single tapping (Fig. 23d). When the surface was tapped by double fingers, about 0.6 V voltage was generated for every single tapping (Fig. 23e). Additionally, only one side of triboelectric pairs were connected. A higher triboelectric effect can be achieved if both sides are connected to the circuit.

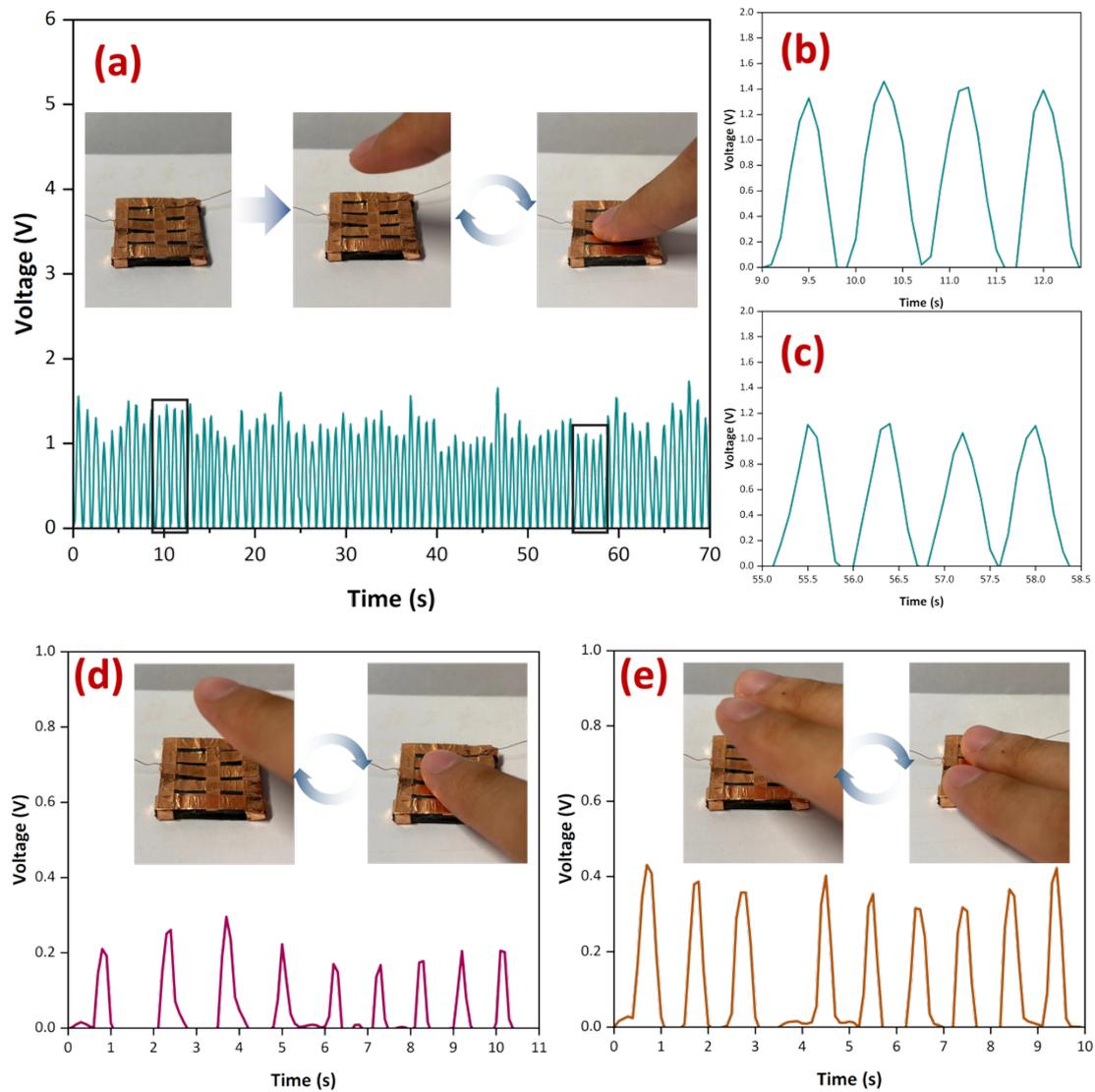

Fig. 23 *(a) Voltage-Time diagram of single finger pressing for 83 cycles in 70 s; (b) four cycles during 9-12.4 s; (c) four cycles during 55-58.5 s; Voltage-Time diagram for (a) single and (b) double finger cyclic tapping*

## 4. Conclusions

An ANF/MXene/SSG sandwich structure has been fabricated. The flexible double layer of ANF/MXene exhibits both high tensile strength, impact resistance, and chemical corrosion of ANF and high conductivity of MXene, exhibiting excellent synergetic effect. SSG/PU composite fabricated in this work exhibit obvious elasticity and shear stiffening effect, overcoming the cold flow phenomenon of pure SSG. The flexible ANF/MXene exhibits tensile strength of 57.15 MPa and tensile strain of over 6%, indicating the strength and toughness modification of pure MXene after the introduction of ANF layer. After the fabrication of the flexible sandwich structure, the A/M/S/M/A structure with a thickness of 4.48mm exhibited an EMI SE of 62.6 dB in the X-band. The absorption attenuation plays the main role in total shielding effectiveness, including microscopic multireflection and ohmic loss absorption. This A/M/S/M/A flexible structure also exhibited the highest impact force attenuation of 71.0%. During impact force attenuation, the SSG core acts as the most important role. Based on the modification of the ANF/MXene/SSG sandwich structure, MGN/PDMS TENG exhibited a relatively stable voltage generation. Apart from that, it exhibited voltage of 0.4 V for single-finger tapping and 0.6 V for double-finger tapping. In summary, the ANF/MXene/SSG flexible sandwich structure exhibits excellent EMI shielding and anti-impact performance. After modification, the MGN/PDMS TENG can generate a voltage signal under finger pressure, indicating the potential for human protection and human body signal detection as a wearable device.

**References**

[1] X.-Y. Wang, S.-Y. Liao, Y.-J. Wan, P.-L. Zhu, Y.-G. Hu, T. Zhao *et al.*, "Electromagnetic interference shielding materials: recent progress, structure design, and future perspective", Journal of Materials Chemistry C, 2022, 10 (1), pp. 44-72.

[2] B. G. Dj and G. J. J. Wessley, "A study on EMI shielding in aircraft: introduction, methods and significance of using electrospun nanocomposites", J. Space Saf. Eng., 2024, 11 (1), pp. 150-160.

[3] J. Kruzelak, A. Kvasnicakova, K. Hlozekova and I. Hudec, "Progress in polymers and polymer composites used as efficient materials for EMI shielding", Nanoscale Adv., 2021, 3 (1), pp. 123-172.

[4] V. Shukla, "Review of electromagnetic interference shielding materials fabricated by iron ingredients", Nanoscale Adv., 2019, 1 (5), pp. 1640-1671.

[5] C. Liang, H. Qiu, Y. Zhang, Y. Liu and J. Gu, "External field-assisted techniques for polymer matrix composites with electromagnetic interference shielding", Sci Bull (Beijing), 2023, 68 (17), pp. 1938-1953.

[6] Q. Zhang, Q. Wang, J. Cui, S. Zhao, G. Zhang, A. Gao *et al.*, "Structural design and preparation of Ti(3)C(2)T(x) MXene/polymer composites for absorption-dominated electromagnetic interference shielding", Nanoscale Adv., 2023, 5 (14), pp. 3549-3574.

[7] X. Wang, Y. Tao, C. Zhao, M. Sang, J. Wu, K. C.-F. Leung *et al.*, "Bionic-leaf vein inspired breathable anti-impact wearable electronics with health monitoring, electromagnetic interference shielding and thermal management", Journal of Materials Science & Technology, 2024, 188, pp. 216-227.

[8] M. Sang, J. Zhang, S. Liu, J. Zhou, Y. Wang, H. Deng *et al.*, "Advanced MXene/shear stiffening composite-based sensor with high-performance electromagnetic interference shielding and anti-impacting Bi-protection properties for smart wearable device", Chem. Eng. J., 2022, 440.

[9] X. Li, M. Liu, Y. Fang, Z. Wu, J. Dong, X. Zhao *et al.*, "Flexible and efficient MXene/PANI/aramid fabrics with high interface durability for wearable electromagnetic wave shielding", Compos. Commun., 2024, 45.

[10] L. Wang, Z. Ma, Y. Zhang, L. Chen, D. Cao and J. Gu, "Polymer-based EMI shielding composites with 3D conductive networks: A mini-review", SusMat, 2021, 1 (3), pp. 413-431.

[11] Y. Bai, B. Zhang, G. Fei and Z. Ma, "Composite polymeric film for stretchable, self-healing, recyclable EMI shielding and Joule heating", Chem. Eng. J., 2023, 478.

[12] T. Chakraborty, S. Saha, K. Gupta, S. Dutta, A. S. Mahapatra, D. Mondal *et al.*, "An effective microwave absorber using multi-layer Design of carbon allotrope based Co$_2$Z hexaferrite-Polymer nanocomposite film for EMI shielding applications", Chem. Eng. J., 2024, 487.

[13] S. Zhang, J. Ye and X. Liu, "Constructing conductive network using 1D and 2D conductive fillers in porous poly(aryl ether nitrile) for EMI shielding", Colloids

Surf., A, 2023, 656.
[14] A. K. Singh, A. Shishkin, T. Koppel and N. Gupta, "Chapter 18 - Porous materials for EMI shielding", in Materials for Potential EMI Shielding Applications, 2020, pp. 287-314.
[15] K. Wang, C. Zhou, Y. Zhou, Y. Fan, J. Chen, L. Hui *et al.*, "Design and research of high-performance electromagnetic interference shielding GO/NiNCs/PMMA microcellular foams", J. Appl. Polym. Sci., 2023, 140 (29).
[16] S.-W. Dai, Y.-L. Gu, L. Zhao, W. Zhang, C.-H. Gao, Y.-X. Wu *et al.*, "Bamboo-inspired mechanically flexible and electrically conductive polydimethylsiloxane foam materials with designed hierarchical pore structures for ultra-sensitive and reliable piezoresistive pressure sensor", Composites, Part B, 2021, 225.
[17] J. Li, L. Wang, H. Luo, Q. Gao, Y. Chen, J. Xiang *et al.*, "Sandwich-like high-efficient EMI shielding materials based on 3D conductive network and porous microfiber skeleton", Colloids Surf., A, 2022, 655.
[18] H. Yahyaei and M. Mohseni, "Nanocomposites Based EMI Shielding Materials", in Advanced Materials for Electromagnetic Shielding, 2018, pp. 263-288.
[19] A. A. Ramachandran and S. Thomas, "Polymer Nanocomposites for EMI Shielding Application", in Polymer Nanocomposite Materials, 2021, pp. 267-284.
[20] M. D. Teli and S. P. Valia, "Graphene and CNT Based EMI Shielding Materials", in Advanced Materials for Electromagnetic Shielding, 2018, pp. 241-261.
[21] S. Parmar, B. Ray, A. Ashok, U. Shilpa and S. Datar, "Highly tunable rGO composites for EMI shielding inks: A tailored approach to modulate broadband EMI shielding", Synth. Met., 2024, 301.
[22] M. H. Al-Saleh and H. M. El-Methaly, "Influence of PLA/HDPE Ratio and CNT content on the morphology, electrical, and EMI shielding of CNT-filled PLA/HDPE blends", Synth. Met., 2024, 304.
[23] Q. Gao, J. Qin, B. Guo, X. Fan, F. Wang, Y. Zhang *et al.*, "High-performance electromagnetic interference shielding epoxy/Ag nanowire/thermal annealed graphene aerogel composite with bicontinuous three-dimensional conductive skeleton", Composites, Part A, 2021, 151.
[24] S. Piao, Z. Jiang, S. Li, T. Park, Y. Kim, E. Lee *et al.*, "Recycled polyethylene terephthalate/ FeCo@C/ Silver nanowire/ polyimide sandwich membrane for electrothermal heating and electromagnetic interference shielding", Journal of Alloys and Compounds, 2024, 987.
[25] Q. Gao, G. Zhang, Y. Zhang, X. Fan, Z. Wang, S. Zhang *et al.*, "Absorption dominated high-performance electromagnetic interference shielding epoxy/functionalized reduced graphene oxide/Ni-chains microcellular foam with asymmetric conductive structure", Composites Science and Technology, 2022, 223.
[26] Z. Zhang, L. Hu, R. Wang, S. Zhang, L. Fu, M. Li *et al.*, "Advances in Monte Carlo Method for Simulating the Electrical Percolation Behavior of Conductive Polymer Composites with a Carbon-Based Filling", Polymers (Basel), 2024, 16 (4).
[27] H. Yuan, H. Chen, M. Li, L. Liu and Z. Liu, "Percolation threshold and electrical

conductivity of conductive polymer composites filled with curved fibers in two-dimensional space", Soft Matter, 2023, 19 (37), pp. 7149-7160.

[28] D. Hu, X. Huang, S. Li and P. Jiang, "Flexible and durable cellulose/MXene nanocomposite paper for efficient electromagnetic interference shielding", Composites Science and Technology, 2020, 188.

[29] Z. Ma, S. Kang, J. Ma, L. Shao, Y. Zhang, C. Liu *et al.*, "Ultraflexible and Mechanically Strong Double-Layered Aramid Nanofiber–Ti3C2Tx MXene/Silver Nanowire Nanocomposite Papers for High-Performance Electromagnetic Interference Shielding", ACS Nano, 2020, 14 (7), pp. 8368-8382.

[30] Y.-J. Wan, P.-L. Zhu, S.-H. Yu, R. Sun, C.-P. Wong and W.-H. Liao, "Graphene paper for exceptional EMI shielding performance using large-sized graphene oxide sheets and doping strategy", Carbon, 2017, 122, pp. 74-81.

[31] S. Zhang, H. Sun, T. Lan, Z. Bai and X. Liu, "Facile preparation of graphene film and sandwiched flexible poly(arylene ether nitrile)/graphene composite films with high EMI shielding efficiency", Composites, Part A, 2022, 154, pp. 106777.

[32] A. VahidMohammadi, J. Rosen and Y. Gogotsi, "The world of two-dimensional carbides and nitrides (MXenes)", Science, 2021, 372 (6547).

[33] X. Zhan, C. Si, J. Zhou and Z. Sun, "MXene and MXene-based composites: synthesis, properties and environment-related applications", Nanoscale Horiz., 2020, 5 (2), pp. 235-258.

[34] "MXenes as EMI Shielding Materials", in Two-Dimensional Materials for Electromagnetic Shielding, 2021, pp. 125-176.

[35] M. R. Lukatskaya, O. Mashtalir, C. E. Ren, Y. Dall'Agnese, P. Rozier, P. L. Taberna *et al.*, "Cation Intercalation and High Volumetric Capacitance of Two-Dimensional Titanium Carbide", Science, 2013, 341 (6153), pp. 1502-1505.

[36] F. Shahzad, M. Alhabeb, C. B. Hatter, B. Anasori, S. Man Hong, C. M. Koo *et al.*, "Electromagnetic interference shielding with 2D transition metal carbides (MXenes)", Science, 2016, 353 (6304), pp. 1137-1140.

[37] M. Boota, B. Anasori, C. Voigt, M. Q. Zhao, M. W. Barsoum and Y. Gogotsi, "Pseudocapacitive Electrodes Produced by Oxidant-Free Polymerization of Pyrrole between the Layers of 2D Titanium Carbide (MXene)", Adv. Mater., 2016, 28 (7), pp. 1517-1522.

[38] R. Liu, M. Miao, Y. Li, J. Zhang, S. Cao and X. Feng, "Ultrathin Biomimetic Polymeric Ti3C2Tx MXene Composite Films for Electromagnetic Interference Shielding", ACS Appl. Mater. Interfaces, 2018, 10 (51), pp. 44787-44795.

[39] W.-T. Cao, F.-F. Chen, Y.-J. Zhu, Y.-G. Zhang, Y.-Y. Jiang, M.-G. Ma *et al.*, "Binary Strengthening and Toughening of MXene/Cellulose Nanofiber Composite Paper with Nacre-Inspired Structure and Superior Electromagnetic Interference Shielding Properties", ACS Nano, 2018, 12 (5), pp. 4583-4593.

[40] M. Yang, K. Cao, L. Sui, Y. Qi, J. Zhu, A. Waas *et al.*, "Dispersions of Aramid Nanofibers: A New Nanoscale Building Block", ACS Nano, 2011, 5 (9), pp. 6945-6954.

[41] B. Yang, L. Wang, M. Zhang, J. Luo and X. Ding, "Timesaving, High-Efficiency Approaches To Fabricate Aramid Nanofibers", ACS Nano, 2019, 13 (7), pp.

7886-7897.

[42] W. Tian, T. Qiu, Y. Shi, L. He and X. Tuo, "The facile preparation of aramid insulation paper from the bottom-up nanofiber synthesis", Mater. Lett., 2017, 202, pp. 158-161.

[43] J. Zhou, S. Wang, J. Zhang, Y. Wang, H. Deng, S. Sun et al., "Enhancing Bioinspired Aramid Nanofiber Networks by Interfacial Hydrogen Bonds for Multiprotection under an Extreme Environment", ACS Nano, 2023, 17 (4), pp. 3620-3631.

[44] C. Zhao, X. Gong, S. Wang, W. Jiang and S. Xuan, "Shear Stiffening Gels for Intelligent Anti-impact Applications", Cell Rep. Phys. Sci., 2020, 1 (12).

[45] C. Clavaud, A. Berut, B. Metzger and Y. Forterre, "Revealing the frictional transition in shear-thickening suspensions", Proc. Natl. Acad. Sci. U. S. A., 2017, 114 (20), pp. 5147-5152.

[46] S. Wang, W. Jiang, W. Jiang, F. Ye, Y. Mao, S. Xuan et al., "Multifunctional polymer composite with excellent shear stiffening performance and magnetorheological effect", J. Mater. Chem. C, 2014, 2 (34), pp. 7133-7140.

[47] W. Jiang, X. Gong, S. Wang, Q. Chen, H. Zhou, W. Jiang et al., "Strain rate-induced phase transitions in an impact-hardening polymer composite", Appl. Phys. Lett., 2014, 104 (12).

[48] C. Xu, Y. Wang, J. Wu, S. Song, S. Cao, S. Xuan et al., "Anti-impact response of Kevlar sandwich structure with silly putty core", Composites Science and Technology, 2017, 153, pp. 168-177.

[49] Y. Wang, L. Ding, C. Zhao, S. Wang, S. Xuan, H. Jiang et al., "A novel magnetorheological shear-stiffening elastomer with self-healing ability", Composites Science and Technology, 2018, 168, pp. 303-311.

[50] Q. Wu, H. Xiong, Y. Peng, Y. Yang, J. Kang, G. Huang et al., "Highly Stretchable and Self-Healing "Solid–Liquid" Elastomer with Strain-Rate Sensing Capability", ACS Appl. Mater. Interfaces, 2019, 11 (21), pp. 19534-19540.

[51] S. Wang, S. Xuan, Y. Wang, C. Xu, Y. Mao, M. Liu et al., "Stretchable Polyurethane Sponge Scaffold Strengthened Shear Stiffening Polymer and Its Enhanced Safeguarding Performance", ACS Appl. Mater. Interfaces, 2016, 8 (7), pp. 4946-4954.

[52] F.-R. Fan, Z.-Q. Tian and Z. Lin Wang, "Flexible triboelectric generator", Nano Energy, 2012, 1 (2), pp. 328-334.

[53] M. Ma, Z. Kang, Q. Liao, Q. Zhang, F. Gao, X. Zhao et al., "Development, applications, and future directions of triboelectric nanogenerators", Nano Research, 2018, 11 (6), pp. 2951-2969.

[54] D. Choi, Y. Lee, Z. H. Lin, S. Cho, M. Kim, C. K. Ao et al., "Recent Advances in Triboelectric Nanogenerators: From Technological Progress to Commercial Applications", ACS Nano, 2023, 17 (12), pp. 11087-11219.

[55] S. Bairagi, I. Shahid ul, C. Kumar, A. Babu, A. K. Aliyana, G. Stylios et al., "Wearable nanocomposite textile-based piezoelectric and triboelectric nanogenerators: Progress and perspectives", Nano Energy, 2023, 118.

[56] J. Shao, T. Jiang and Z. Wang, "Theoretical foundations of triboelectric

nanogenerators (TENGs)", Sci. China Technol. Sci., 2020, 63 (7), pp. 1087-1109.

[57] Z. L. Wang and A. C. Wang, "On the origin of contact-electrification", Mater. Today, 2019, 30, pp. 34-51.

[58] H. Zhang, L. Yao, L. Quan and X. Zheng, "Theories for triboelectric nanogenerators: A comprehensive review", Nanotechnol. Rev., 2020, 9 (1), pp. 610-625.

[59] W. Wang, J. Zhou, S. Wang, F. Yuan, S. Liu, J. Zhang et al., "Enhanced Kevlar-based triboelectric nanogenerator with anti-impact and sensing performance towards wireless alarm system", Nano Energy, 2022, 91.

[60] K.-X. Hou, X. Dai, S.-P. Zhao, L.-B. Huang and C.-H. Li, "A damage-tolerant, self-healing and multifunctional triboelectric nanogenerator", Nano Energy, 2023, 116.

[61] S. Thukral, S. Kovac and M. Paturu, "Chapter 29 - t-test", in Translational Interventional Radiology, 2023, pp. 139-143.

[62] S. C. Ray, "Chapter 2 - Application and Uses of Graphene Oxide and Reduced Graphene Oxide", in Applications of Graphene and Graphene-Oxide Based Nanomaterials, 2015, pp. 39-55.

[63] M. K. Aswathi, A. V. Rane, A. R. Ajitha, S. Thomas and M. Jaroszewski, "EMI Shielding Fundamentals", in Advanced Materials for Electromagnetic Shielding, 2018, pp. 1-9.

[64] L. Liang, P. Xu, Y. Wang, Y. Shang, J. Ma, F. Su et al., "Flexible polyvinylidene fluoride film with alternating oriented graphene/Ni nanochains for electromagnetic interference shielding and thermal management", Chem. Eng. J., 2020, 395, pp. 125209.

[65] S. Ramachandra Rao, "Chapter 3 - Physical and Physico-Chemical Processes", in Waste Management Series, 2006, pp. 35-69.

[66] C. Xie, L. He, Y. Shi, Z.-X. Guo, T. Qiu and X. Tuo, "From Monomers to a Lasagna-like Aerogel Monolith: An Assembling Strategy for Aramid Nanofibers", ACS Nano, 2019, 13 (7), pp. 7811-7824.

[67] S.-W. Kim, J.-K. Kim, S. Jung, J. W. Lee, C. Yang and J. M. Baik, "Choice of Materials for Triboelectric Nanogenerators", in Handbook of Triboelectric Nanogenerators, 2023, Chapter 13-1, pp. 1-50.

[68] Y.-H. Ji, Y. Liu, Y.-Q. Li, H.-M. Xiao, S.-S. Du, J.-Y. Zhang et al., "Significantly enhanced electrical conductivity of silver nanowire/polyurethane composites via graphene oxide as novel dispersant", Composites Science and Technology, 2016, 132, pp. 57-67.